\pdfoutput=1
\PassOptionsToPackage{table}{xcolor}

\documentclass[sigconf, nonacm]{acmart}

\setcopyright{none}
\acmISBN{978-x-xxxx-xxxx-x/YY/MM}
\acmDOI{XXXXXXX.XXXXXXX}
\usepackage{array}

\newcommand{\fs}{FoldSim}
\newcommand{\fsmail}[1]{\href{mailto:#1@ucla.edu}{#1}}   %
\newcommand{\ns}{\,ns}
\newcommand{\us}{\,\textmu{}s}
\newcommand{\mhz}{\,MHz}

\graphicspath{{figs/}}

\newcommand{\ignore}[1]{}  

\begin{document}

\setlength{\parskip}{0pt plus 1.5pt}
\makeatletter
\def\@textbottom{\vskip \z@ \@plus 8\p@}
\makeatother

\title{FoldSim: Scaling Cycle-Exact RTL Simulation with Hierarchical Folding and Static Scheduling
}

\author{Chengxuan Wang, Nikhil Rout, Yang-gon Kim, Jacob Leigh, Yufan Du, Blaise Tine}
\affiliation{%
  \institution{University of California, Los Angeles}
  \country{%
    \{\fsmail{chengxuan99}, \fsmail{nikhilrout97}, \fsmail{yanggon},
    \fsmail{jlaurenceleigh}, \fsmail{yufandu}\}@ucla.edu,
    \href{mailto:blaisetine@cs.ucla.edu}{blaisetine@cs.ucla.edu}}}
\renewcommand{\shortauthors}{Chengxuan Wang, Nikhil Rout, Yang-gon Kim, Jacob Leigh, Yufan Du, Blaise Tine}

\begin{abstract}
RTL simulation is essential for hardware design and verification, yet its execution time grows rapidly with design scale and complexity. CPU- and GPU-based simulators exploit parallelism to accelerate RTL evaluation, but achieving high simulation throughput for large designs remains challenging. FPGA-based simulators offer higher throughput by mapping RTL computations directly onto hardware, but their scalability is constrained by FPGA resource capacity.

We present FoldSim, an automated framework that enables cycle-exact simulation of large RTL designs on a single FPGA through hierarchical folding. FoldSim maps instances across the design hierarchy onto shared logic while maintaining their independent state. Given foldable module classes, its compiler constructs a static schedule that coordinates their execution, reducing dynamic synchronization overhead and making state accesses predictable. FoldSim backs state that exceeds on-chip capacity with external memory and brings it to the shared circuits as needed. A configurable DRAM model separates target-memory timing from host delays. The compiler explores replica allocations and state placement to balance resource savings against simulation time.

On one AMD Alveo V80, FoldSim runs an 8-SM GPU
and a 1,024-node Rocket datacenter with 4,096
processor cores. The unfolded implementations require
approximately 16$\times$ and 200$\times$ as many LUTs
as FoldSim's, respectively. The datacenter
keeps about 1.9\,GB of state in HBM, while a 4-SM
GPU completes a $1024\times1024$ GEMM in 4.24
seconds, versus 26 hours in software RTL simulation.

\end{abstract}

\maketitle
\begingroup\renewcommand\thefootnote{}\footnotetext{\fs{} is open source at \url{https://github.com/orcaslab/foldsim}.}\endgroup

\section{Introduction}
\label{sec:intro}

\begin{figure}[t]
  \centering
  \includegraphics[width=0.99\columnwidth]{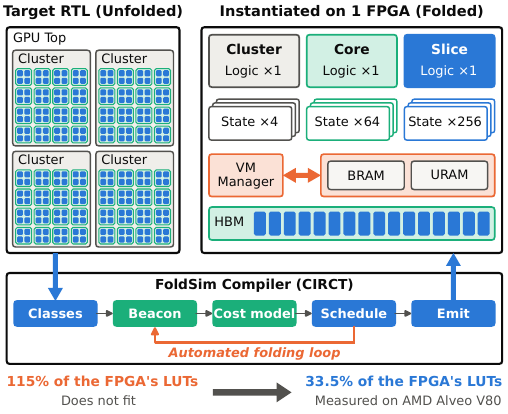}
  \caption{
  FoldSim Framework Overview.
  }
  \label{fig:framework}
\end{figure}

RTL simulation plays a critical role in hardware development, enabling designers to verify implementations and evaluate architectural behavior before fabrication. 
Advancing a simulated clock cycle requires evaluating logic dependencies and updating design state, with the computational effort increasing as designs grow in scale and complexity. 
Combined with the millions or billions of cycles often needed to exercise realistic workloads, this makes RTL simulation a major bottleneck in hardware verification and design exploration.

Prior work accelerates RTL simulation by exploiting parallelism. RepCut~\cite{wang2023repcut} partitions circuit evaluation across CPU threads; GEM~\cite{guo2025gem} and RTLflow~\cite{lin2022rtlflow} exploit GPUs. Manticore~\cite{emami2023manticore} and Parendi~\cite{emami2025parendi} use specialized parallel processors, while LOTUS~\cite{elsabbagh2026lotus} executes simulation tasks on lightweight processors across multiple FPGAs. These approaches reduce execution time but still incur instruction-execution and communication overhead. Direct FPGA mapping instead implements RTL as circuits, offering high throughput when the design fits on the device.

 Scaling FPGA simulation by adding devices introduces another constraint. FireSim~\cite{karandikar2018firesim} demonstrated a 1,024-node datacenter simulation across 256 FPGAs, but each FPGA-hosted RTL model must still fit its device. FireAxe~\cite{whangbo2024fireaxe} partitions designs across multiple FPGAs to overcome this limit, at the cost of cross-device communication. Such communication can slow simulation when dependencies cross partitions within a target cycle, while deployments spanning servers introduce further communication overhead. Keeping a tightly coupled design on one FPGA would avoid these costs, provided its hardware footprint could be reduced.

GPU RTL makes this fitting problem especially acute. A Vortex-based GPU~\cite{tine2021vortex} with A100-like execution resources~\cite{nvidia2020ampere} requires an estimated 71\% of an Alveo V80's total LUT capacity for a single unfolded SM alone. Once the FPGA shell and memory simulator are included, even one SM exceeds the available resources. Partitioning its tightly coupled logic across multiple FPGAs risks introducing frequent, latency-sensitive communication over inter-device links. Instead, GPU SMs contain repeated processing blocks, tensor-core lanes, and floating-point units. Time-multiplexing these structures can reduce the hardware footprint without partitioning the SM across FPGAs. The FAME taxonomy~\cite{tan2010fame} describes time-multiplexed simulation, while Golden Gate~\cite{magyar2019goldengate} and later instance-threading work~\cite{magyar2022thesis} share logic among identical RTL instances. We call a module together with all of its identical instances a \emph{class}. However, folding many classes across multiple hierarchy levels must coordinate dependent modules without excessive synchronization, balance sharing against throughput, and preserve per-instance state that may exceed on-chip memory capacity.

We present \fs{}, which addresses these challenges through hierarchical folding, compile-time scheduling of inter-class dependencies, and state virtualization. Its compiler explores replica allocations and state placement to balance FPGA capacity against simulation throughput (Fig.~\ref{fig:framework}). Our contributions are summarized as follows:

\begin{itemize}
    \item \textbf{Static scheduling.}
    We derive a compile-time execution schedule across folded
    module classes, resolving inter-class dependencies without
    runtime token-driven coordination and reducing synchronization
    overhead and redundant evaluations (Sec.~4.1).

    \item \textbf{Hierarchical folding.}
    We extend instance threading across multiple levels of the RTL
    hierarchy, extracting nested module classes and independently
    selecting physical logic copies for each class to enable
    finer-grained resource sharing (Sec.~4.2).

    \item \textbf{State virtualization.}
    We virtualize per-instance state that folding cannot eliminate,
    using HBM backing storage and schedule-directed transfers to
    scale beyond on-chip storage capacity (Sec.~4.3).

    \item \textbf{Configurable memory modeling.}
    We integrate a configurable DRAM model with folded execution,
    supporting different memory standards and timing parameters
    while decoupling target-memory timing from FPGA memory latency
    (Sec.~4.4).

    \item \textbf{Automated folding.}
    We develop an automated compiler flow that explores replica
    allocations and state placement using static schedules and
    calibrated resource and timing estimates (Sec.~4.5).
\end{itemize}

On a single AMD Alveo V80, FoldSim maps 1,467 RTL instances
across 30 module classes onto just 35 physical copies in a
4-SM Vortex GPU. It also runs a separate 16-slice GPU whose unfolded implementation exceeds the FPGA's logic capacity. For a 1,024-node quad-core Rocket system, FoldSim virtualizes approximately 1.9 GB of state into HBM while executing a bare-metal network workload. Routed comparisons against our token-based folds further show 13--21\% fewer LUTs and 14--21\% higher clock frequency.

\begin{table*}[t]
  \caption{Execution models and scaling tradeoffs in RTL simulation.}
  \label{tab:related}
  \centering\footnotesize
  \setlength{\tabcolsep}{4pt}
  \renewcommand{\arraystretch}{1.10}
  \begin{tabular*}{\textwidth}{@{\extracolsep{\fill}}>{\raggedright\arraybackslash}p{4.3cm}>{\raggedright\arraybackslash}p{4.3cm}>{\raggedright\arraybackslash}p{4.1cm}>{\raggedright\arraybackslash}p{3.8cm}@{}}
  \toprule
  \textbf{System} & \textbf{Execution model} &
  \textbf{Scaling approach} & \textbf{Performance cost} \\
  \midrule
  RepCut / RTLflow / GEM~\cite{wang2023repcut,lin2022rtlflow,guo2025gem}
  & Compiled RTL computations on CPU/GPU
  & Partition, stimulus parallelism, GPU
  & Instruction, data transfer, sync \\
  Manticore / Parendi / LOTUS~\cite{emami2023manticore,emami2025parendi,elsabbagh2026lotus}
  & Compiled tasks on FPGA or IPU cores
  & Static BSP, IPU scale, task dataflow
  & Task execution, inter-core traffic \\
  \midrule
  FireSim~\cite{karandikar2018firesim}
  & Spatial RTL circuits
  & RTL replication; multi-FPGA nodes
  & Host/memory waits; FPGA capacity\\
  FireAxe~\cite{whangbo2024fireaxe}
  & RTL circuits distributed across FPGAs
  & Aggregate capacity of multiple FPGAs
  & Inter-FPGA link latency \\
  \midrule
  Golden Gate~\cite{magyar2019goldengate,magyar2022thesis}
  & Time-shared RTL module logic
  & Share logic; retain per-instance state
  & Instance visits; inter-model tokens \\
  \midrule
  FoldSim
  & Hierarchically shared RTL classes
  & Per-class copies; HBM-backed state
  & Scheduled II and state/memory stalls \\
  \bottomrule
  \end{tabular*}
\end{table*}

\section{Background}
\label{sec:background}\label{sec:bg-fame}

An FPGA-accelerated simulator executes a target design on a host FPGA.
A \emph{target cycle} is one clock cycle of the simulated design; a
\emph{host cycle} is one FPGA clock cycle. A simulator may require
several host cycles to execute one target cycle.

\begin{figure}[t]
  \centering
  \setlength{\unitlength}{\dimexpr\columnwidth/68\relax}%
  \begin{picture}(68,29.6)
    \put(0,0){\includegraphics[width=\columnwidth]{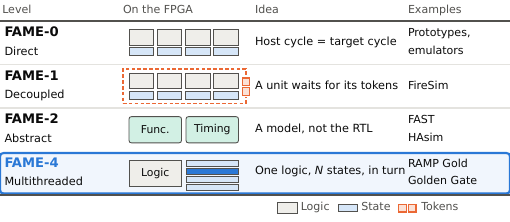}}
    \sffamily\fontsize{5.3}{6}\selectfont
    \put(60.2,17.58){\cite{karandikar2018firesim}}
    \put(58.4,12.93){\cite{chiou2007fast}}
    \put(59.7,10.63){\cite{pellauer2011hasim}}
    \put(62.4,7.13){\cite{tan2010rampgold}}
    \put(63.3,4.83){\cite{magyar2022thesis}}
  \end{picture}
  \caption{The FAME taxonomy and its three techniques.}
  \label{fig:fame}
\end{figure}

A \emph{direct implementation} (FAME-0) maps each RTL module to hardware,
advancing one target cycle per host cycle, like an FPGA prototype.
The design must fit on the FPGA, and memory and I/O follow host timing. The FAME taxonomy~\cite{tan2010fame}
identifies three combinable techniques (Fig.~\ref{fig:fame}).
\emph{Decoupling} (FAME-1) separates target time from host cycles:
components advance when their required input \emph{tokens} arrive.
A target cycle may span several host cycles to accommodate memory
and I/O delays without changing target-visible timing.
\emph{Abstraction} (FAME-2) replaces RTL with a functional or timing
model. Such models can be cycle-accurate but cannot expose bugs in
the RTL implementation itself.
Here, \emph{cycle-exact} simulation preserves the RTL's state transitions
and observable behavior at each target-cycle boundary.

A \emph{multithreaded} simulator (FAME-4) shares logic among identical
modules while retaining their independent state. A module's logic
computes its outputs and next state from its inputs and current state,
held in registers and memories. For $N$ identical instances, the
simulator builds one logic copy but preserves $N$ separate states.
In successive host cycles, the copy loads each instance's state,
evaluates its logic, and stores the result. This RTL multithreading,
called \emph{folding}, requires at least $N$ host cycles per target cycle.
With $k$ logic copies, the lower bound becomes $\lceil N/k \rceil$
host cycles, trading FPGA logic capacity for simulation throughput.

The three FAME techniques are composable rather than mutually
exclusive. FAME-3 (1+2) combines decoupling and
abstraction, as in HAsim~\cite{pellauer2011hasim}; FAME-7 (1+2+4) also
adds multithreading, as in RAMP Gold~\cite{tan2010rampgold}.
FireSim~\cite{karandikar2018firesim} initially used FAME-1, while
Golden Gate~\cite{magyar2022thesis} enabled FAME-5 (1+4) through RTL
instance threading. FireAxe~\cite{whangbo2024fireaxe} can use FAME-5
within multi-FPGA partitions. FoldSim is also FAME-5, extending RTL
folding with hierarchical sharing, static scheduling, and HBM-backed
state virtualization.

\section{Related Work}
\label{sec:related}
\providecommand{\eqinfig}[1]{\refstepcounter{equation}\label{#1}}

\subsection{Programmable and circuit-based simulation}
\label{sec:other}
RTL simulators exploit parallelism through different execution substrates. RepCut~\cite{wang2023repcut} reduces synchronization between CPU threads by replicating logic at partition boundaries. RTLflow~\cite{lin2022rtlflow} exploits parallel verification stimuli, while GEM~\cite{guo2025gem} maps circuit operations onto a GPU-executed virtual VLIW engine. Manticore~\cite{emami2023manticore} employs statically scheduled execution on an FPGA manycore, and Parendi~\cite{emami2025parendi} distributes simulation across Graphcore IPU cores. LOTUS~\cite{elsabbagh2026lotus} further specializes task execution with over 2,000 lightweight RISC-V processors, task-dataflow scheduling, and dedicated control hardware distributed across eight FPGAs. These approaches reduce synchronization but still incur instruction-execution and data-movement overhead. FPGA-based simulators can instead realize RTL computations directly as circuits, at the cost of limited physical capacity.

\begin{figure}[t]
  \centering
  \setlength{\unitlength}{\dimexpr\columnwidth/68\relax}%
  \begin{picture}(0,0)\def\HyperRaiseLinkDefault{0pt}%
    \put(0,9.4){\eqinfig{eq:unfolded}}\put(0,5.4){\eqinfig{eq:tcycle}}%
    \put(22.1,9.4){\eqinfig{eq:fpgas}}\put(22.1,5.4){\eqinfig{eq:link}}%
    \put(44.2,9.4){\eqinfig{eq:folded}}\put(44.2,5.4){\eqinfig{eq:hbound}}%
  \end{picture}%
  \includegraphics[width=\columnwidth]{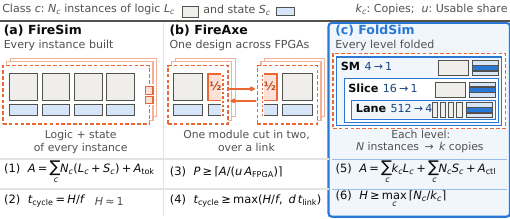}
  \caption{Illustrative area and timing tradeoffs for (a) spatial
  replication, (b) partitioning, and (c) hierarchical folding.}
  \Description{Three schematic FPGA mappings: spatial replication builds
  every instance; partitioning distributes a design across FPGA links;
  hierarchical folding shares logic across repeated classes while retaining
  independent instance state. Six equations show simplified cost bounds.}
  \label{fig:place}
\end{figure}

\subsection{Scaling spatial mappings}
\label{sec:bg-firesim}\label{sec:bg-fireaxe}
In an unfolded mapping (Fig.~\ref{fig:place}(a)), each RTL instance retains separate logic and state. This avoids the host-cycle overhead of sharing logic but requires the full design to fit the available resources. Eqs.~(\hyperref[eq:unfolded]{1})--(\hyperref[eq:tcycle]{2}) illustrate the area and target-cycle time $H/f$, including token-control overhead $A_{\mathrm{tok}}$ for decoupled simulation. The depicted $H\approx1$ is an ideal direct-mapping case, not a guarantee for FireSim. FireSim~\cite{karandikar2018firesim} demonstrated 1,024 quad-core nodes across 256 FPGAs, but each FPGA-hosted model must fit its allocated device. Figure~\ref{fig:place}(a) illustrates its unfolded spatial baseline, not the full set of later FireSim optimizations.

FireAxe~\cite{whangbo2024fireaxe} overcomes this per-device limit by partitioning individual RTL designs across FPGAs (Fig.~\ref{fig:place}(b)). Its exact mode preserves target-cycle behavior even across combinational partition boundaries, although dependent token exchanges introduce communication latency. FireAxe also uses FAME-5 threading to amortize this latency. Eq.~(\hyperref[eq:fpgas]{3}) is an aggregate-area lower bound on FPGA count; Eq.~(\hyperref[eq:link]{4}) illustrates a latency bound for a critical path with $d$ sequential link transfers, not transfers hidden by overlap. Partitioning therefore trades added hardware capacity for communication whose cost depends on the cut and its dependencies.

\subsection{Folding repeated structure}
\label{sec:potential}\label{sec:rw-goldengate}
Folding shares module logic among instances while retaining independent state (Fig.~\ref{fig:place}(c)). With $k_c$ copies for $N_c$ instances of class $c$, logic replication scales with $k_c$ and logical state with $N_c$ (Eq.~(\hyperref[eq:folded]{5})); HBM-backed state consumes different FPGA resources. A target cycle requires at least $\max_c\lceil N_c/k_c\rceil$ host cycles (Eq.~(\hyperref[eq:hbound]{6})), excluding coordination and memory stalls. In the evaluated four-SM GPU, 16 unfolded compute slices alone require an estimated 2.96\,M LUTs, exceeding the V80's 2.57\,M.

RAMP Gold~\cite{tan2010rampgold} and HAsim~\cite{pellauer2011hasim} explored time-multiplexed abstract hardware models. Golden Gate~\cite{magyar2019goldengate} and subsequent work~\cite{magyar2022thesis} automated RTL instance threading; FireAxe~\cite{whangbo2024fireaxe} can also apply it within partitions. FireSim's FAME-5 pass threads one annotated module class into one physical copy. Nested submodules are transformed but not independently folded: a threaded SM still contains four processing blocks. Static round-robin scheduling selects instances \emph{within} the model, while token handshakes coordinate \emph{between} models; unavailable inputs can waste slots. The pass retains thread-indexed state on chip rather than virtualizing it in HBM.

These limits matter when a GPU must fit on one FPGA. A single unfolded SM requires an estimated 1.82\,M LUTs (71\% of the V80's total), before adding the FPGA shell and memory simulator. Threading SMs retains the full SM; threading a smaller class leaves other repeated logic intact. FoldSim instead assigns independent copy counts to classes throughout the hierarchy (Sec.~4.2), orders their dependencies at compile time rather than through inter-class token handshakes (Sec.~4.1), and backs excess instance state with HBM (Sec.~4.3).

\section{Methodology}
\label{sec:design}

FoldSim transforms repeated instances of the same RTL module into shared
logic copies while preserving the independent state of each
instance (Fig.~\ref{fig:framework}). We call a module together with all of
its identical instances a \emph{class}; a class has $N$ instances and
$k$ logic copies, each copy serving $N/k$ of them. Realizing this transformation requires efficiently coordinating instance execution and organizing logic sharing across multiple levels of the RTL hierarchy. 
We first introduce a compile-time schedule that coordinates instance execution across folded modules (Sec.~\ref{sec:static}). We then describe how FoldSim organizes and shares logic across the design hierarchy (Sec.~\ref{sec:hier}). The subsequent sections address scalable state virtualization (Sec.~\ref{sec:virt}), configurable memory modeling (Sec.~\ref{sec:mem}), and automated folding (Sec.~\ref{sec:flow}).

\subsection{Static scheduling}
\label{sec:static}\label{sec:contract}\label{sec:frames}

\begin{figure}[t]
  \centering
  \includegraphics[width=\columnwidth]{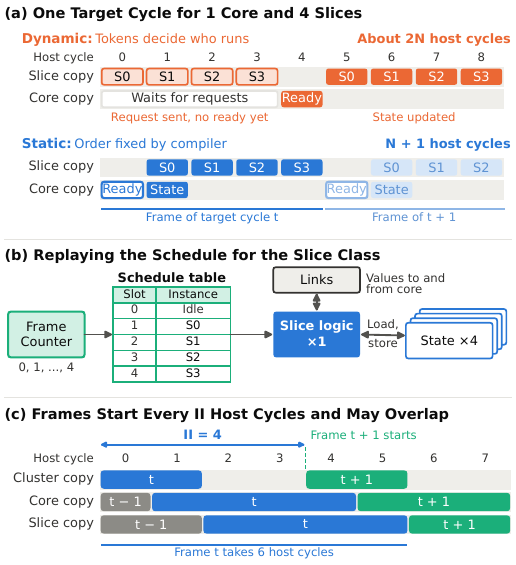}
  \caption{Static scheduling of a folded design.} 
  
  \Description{Two timelines of host cycles and a block diagram. With
  tokens, the slice copy visits its four slices twice and the target cycle
  takes about 2N host cycles. With a static schedule, the core's ready pass
  runs first and each slice runs once, for N+1 host cycles. A frame counter
  indexes a schedule table that selects the instance whose state the shared
  slice logic loads and stores. A third timeline shows the cluster, core
  and slice copies over eight host cycles, with frames that start four host
  cycles apart and each last six.}
  \label{fig:static}\label{fig:tokens}\label{fig:overlap}
\end{figure}

A folded simulator must decide which instance each logic copy evaluates in
every host cycle. We explain how FoldSim decides on a small target, a core
communicating with four functional unit slices that share one logic copy
(Fig.~\ref{fig:static}a). In
every target cycle each slice sends a request to the core, and the core
answers with a ready signal that the slices need in that same cycle.

\subsubsection{Why we fix the order at compile time}
Prior work on folding relies on tokens to schedule work on each class. An
instance runs once the tokens of all its inputs have arrived, so the order
of execution is determined at run time. In the example of
Fig.~\ref{fig:static}a, the slices wait for ready and the core waits for
their requests, so the slice copy visits each of its four instances twice.
A target cycle takes about $2N$ host cycles, while the shared copy needs
only $N$ (Fig.~\ref{fig:static}a, top). Yet nothing in this exchange
depends on data. Which module waits for which is set by how the modules are
connected in the RTL, not by the values they exchange, so it is the same
in every target cycle. FoldSim therefore works the order out statically
once in the compiler, and the simulator only replays it.

\subsubsection{What fixes the order}
The compiler classifies every output of every folded module. An output is
\emph{early} if it depends only on the module's own state, so no consumer
waits for it. An
output is \emph{late} if it depends combinationally on an input of the same
cycle. In the example the requests are early and ready is late. The order
is constrained only by late outputs, because the module that reads one has
to wait for its producer to run first. Here, the slices cannot run until
the core has produced ready.

FoldSim meets these constraints with two kinds of \emph{passes}. Every
instance has one \emph{state pass}, which writes its next state. A class
with late outputs may also receive a \emph{ready pass}, which evaluates the
same logic but keeps only the late outputs. In the example, the core's
ready pass runs first and computes ready. Each slice then runs once with
ready already available, so the target cycle takes $N+1$ host cycles: one
for the ready pass and one per slice (Fig.~\ref{fig:static}a, bottom).

\subsubsection{Replaying the schedule}
The compiler's schedule for one target cycle is a \emph{frame}, and each
host cycle within it is a \emph{slot}. Replaying a frame takes only a
counter and a table (Fig.~\ref{fig:static}b). The counter advances one slot
per host cycle, and the table, written by the compiler, tells the logic
copy which instance to evaluate in each slot. In every slot the copy loads
that instance's state, evaluates it, and writes back the new state. Because
the order is fixed in advance, the copy never waits for tokens or checks
whether an instance is ready to run.

\subsubsection{Computing the schedule}
A new frame starts every II host cycles, where II is the \emph{initiation
interval}. Within every II host cycles, each copy must perform all of its
passes for one target cycle. A class $c$ has $N_c$ instances on $k_c$
copies, so a copy serves $\lceil N_c/k_c \rceil$ instances, and each
instance needs one state pass and $R_c$ ready passes. Hence
\begin{equation}
  \text{II} \;\ge\; \max\Big(2,\; \max_c \Big\lceil \frac{N_c}{k_c}
  \Big\rceil (1 + R_c)\Big).
  \label{eq:ii}
\end{equation}
The most heavily loaded copy sets this bound, and no frame is shorter than
two host cycles. Searching upward from the bound with a list scheduler in
the style of modulo scheduling~\cite{rau1994modulo}, the compiler accepts
the first II at which every pass gets a slot on its copy, after the passes
that produce its inputs.

A frame may last longer than II, because consecutive frames overlap. Each
copy starts the next frame as soon as it has finished its own passes,
without waiting for the other copies to finish theirs. A copy that waits
$x$ host cycles into a frame for its inputs also starts $x$ host cycles
into every later frame, so it still completes one frame of work every II
host cycles. In Fig.~\ref{fig:static}c, a cluster of two cores and four
slices, a frame lasts six host cycles, but a new one starts every four. The
cluster and core copies begin frame $t+1$ while the slice copy is still
finishing frame $t$.

\subsubsection{Why the simulator is exact}
\label{sec:exactarg}%
A static schedule changes only the order in which instances are evaluated
within a frame; every instance still sees its target cycles in the RTL's
order. Each instance gets exactly one state pass per target cycle, plus its
$R_c$ ready passes. By its state pass, every input it reads from other
modules already holds the value it would have in the RTL: early outputs
come from state, and every late output was produced earlier in the frame by
a ready pass. The instance therefore computes the same next state as the
RTL would. The compiler adds no state and no cycle of delay to the target.
The only structure it cannot order is a combinational loop through the
ports of folded modules, where an output depends on itself within the same
cycle; the compiler detects and rejects such designs.

\subsection{Hierarchical folding}
\label{sec:hier}

\begin{figure}[t]
  \centering
  \includegraphics[width=0.97\columnwidth]{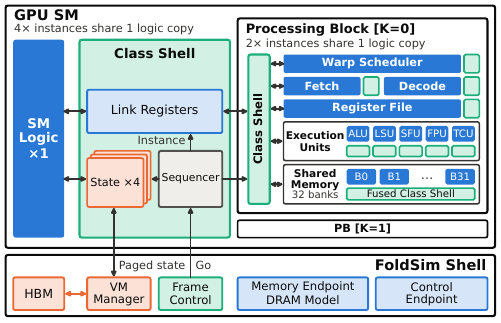}
  \caption{Hierarchical folding of a GPU SM. Each class has its own logic copies
  and a class shell.}
  \Description{A block diagram. A GPU SM box holds one copy of SM logic
  beside a class shell with link registers, a stack of four states and a
  sequencer. To the right are two processing blocks. One is enlarged and
  holds a class shell drawn as one box. Fetch, decode, a warp scheduler, a
  register file and five execution units each carry a small shell. The 32
  banks of a shared memory have no shell of their own and sit on one common
  shell, labelled fused class shell. Two-way arrows connect the class
  shell of the processing block to each of its parts. The other is a small box under
  it. Below,
  the FoldSim shell holds HBM, a VM manager, frame control, a memory
  endpoint with a DRAM model and a control endpoint.}
  \label{fig:hier}
\end{figure}

While static scheduling determines when each instance executes, there is also the question of determining which logic can be shared across instances. Large RTL designs often contain several repeated structures at multiple levels of their hierarchy. Consider the GPU in Fig.~\ref{fig:hier} wich contains multiple Streaming Multiprocessors (SMs). Since each SM has an issue-width of four, there are 4 processing block copies. Going further down the hierarchy, each processing blocks contains a warp scheduler, fetch and decode logic,
register files, multiple execution units, and heavily banked shared memory.

\subsubsection{Why we fold hierarchically}
Folding at only one level leaves substantial replicated logic at other
levels. Here, folding the SMs alone still requires each shared SM logic copy
to contain four processing blocks and all their constituent units.
Conversely, folding only the innermost units leaves their surrounding
logic replicated across instances. FoldSim instead applies folding
throughout the RTL hierarchy, allowing repeated structures at each
level to share physical logic independently.

\subsubsection{The class shell}
FoldSim organizes repeated modules into \emph{classes}, each containing
the logic shared by its instances. To isolate this logic, the compiler
extracts instances of child classes from their parent, and leaves each
class responsible only for its own module logic. Each class is then
equipped with three supporting structures, collectively called the
\emph{class shell}. The \emph{sequencer} replays the class's portion of the static schedule,
selecting the instance to execute in each host cycle. The \emph{state
store} maintains independent state for every instance, while the
\emph{link registers} hold values exchanged with other classes.

\subsubsection{Connecting the levels}
Although classes share physical logic, they retain the parent-child
relationships of the original RTL hierarchy. Communication between
these levels consists of two components: scheduling and data exchange.
For scheduling, each child sequencer receives its control from the
parent sequencer, forming a tree rooted at the frame control of the
FoldSim shell. For data exchange, values pass between classes through
their link registers. Keeping these structures local to each class
avoids wiring that spans multiple levels of the hierarchy.

\subsubsection{Class fusion}

A separate class shell can cost more than the
logic saved when a class is small comparing to the large ones. FoldSim
therefore fuses tightly coupled classes with
corresponding instances into one model, sharing
a shell and eliminating their inter-class link
and scheduling boundary. In Fig.~\ref{fig:hier},
the shared-memory banks use a fused shell instead
of one per bank; Sec.~\ref{sec:flow} details
the fusion criteria.

\subsubsection{Copies per class}
Hierarchical folding also allows each class to have an independent
number of physical logic copies. From Eq.~\ref{eq:ii}, the class with
the greatest number of instances per copy determines the initiation
interval (II), potentially leaving other classes idle for much of
a frame. To balance resource usage and execution throughput,
FoldSim assigns each class $k$ logic copies for its $N$ instances,
ranging from fully folded ($k=1$) to fully unfolded ($k=N$).
Additional copies replicate the class logic but not its per-instance
state. In our example, four SM instances share a single copy of
the SM logic. Within each SM, four processing blocks are mapped
onto two physical logic copies, $K=0$ and $K=1$, with each copy
serving two instances. The selection of ideal $k$ for each class is
discussed later in Sec.~\ref{sec:flow}.

\subsection{State Virtualization}
\label{sec:virt}

While folding reduces the physical logic required for repeated instances,
each instance must still retain its independent state. As the design
scales, this state can exceed the capacity of the on-chip memories
available alongside the shared logic. FoldSim addresses this through
\emph{state virtualization}, exploiting the static schedule to determine when
each instance's state will be accessed. By fetching state ahead of its
scheduled use, FoldSim can store it in larger, slower memories without
altering the target's behavior. State is stored in on-chip block RAM
and UltraRAM, with HBM providing additional capacity when needed.
Here we describe how FoldSim routes state accesses to HBM efficiently and
hides memory latency while preserving cycle-exact execution.

\begin{figure*}[t]
  \centering
  \includegraphics[width=0.492\textwidth]{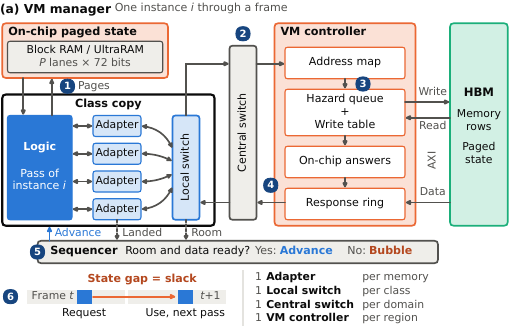}\hfill
  \includegraphics[width=0.492\textwidth]{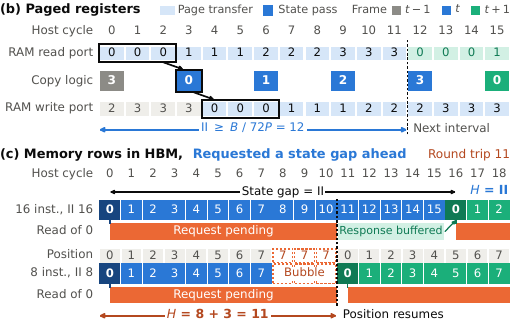}
  \caption{State virtualization of one class copy. Per-instance state lives off the logic copy and is fetched ahead of each pass.}
  \label{fig:statev}\label{fig:vmem}
\end{figure*}

\subsubsection{Why one memory interface per region}
Most states stored in HBM belongs to target memories, such as register
files and cache arrays. Each class may contain several such memories
with different widths and access patterns. For example, the GPU's
operand-reading unit contains four register-file banks with 1,024-bit
rows, while its first-level cache uses 128-bit rows. A single class
evaluation may access all these memories in the same host cycle.

A direct naive implementation would give each memory its own controller
and dedicated HBM interface. However, this requires a wide bus from
every virtual memory instance to the FPGA's HBM interface, with the number of buses
growing alongside the number of memories. These long, wide connections
increase routing congestion and compete with inter-class links in
regions of the FPGA that are already difficult to route. FoldSim
instead consolidates memory traffic within each region through a
shared, hierarchical interface.

\subsubsection{A path that follows the hierarchy}
FoldSim organizes memory access as a tree that follows the class
hierarchy (Fig.~\ref{fig:vmem}a, step~2). At each target memory,
an \emph{adapter} preserves its original interface and row width,
while combining the read and write operations of a host cycle
into a single entry. The example presented contains
four memories, each with its own adapter. Within each class, a \emph{local switch} collects and buffers
entries from its adapters before transmitting them as fixed-size
words, at one word per host cycle. Wider memory rows are transferred
over multiple words, while narrower rows require only one. This
allows memories with different widths and access rates to share
a single bus.

At the next level, a \emph{central switch} combines traffic from
classes belonging to the same placement domain, a group of classes
placed together on the FPGA. Each FPGA region has one \emph{VM
controller} that receives this merged traffic and serves as the
region's sole interface to HBM (step~3). The controller translates
each request's class, memory, instance, and row into an HBM address,
while preserving the ordering of reads and writes to the same row.
Reads to rows that have never been written, or whose writes are
still pending, are resolved on-chip without accessing HBM.
Since HBM responses may arrive out of order, the controller
reorders them to match their requests before returning them
through the tree to the originating adapters (step~4).

Together, the adapters, local switches, central switches, and VM
controllers form the VM manager shown in Fig.~\ref{fig:hier}.
Each connection in this tree is short and registered, with only
one bus leaving each class. Furthermore, each region accesses HBM
through its own ports, avoiding memory traffic across region
boundaries. Since the number of adapters and switches depends on
the number of classes and their memories rather than their instances,
the memory interface does not grow as additional instances are folded.

\subsubsection{Hiding the latency}
Accessing HBM can take tens to hundreds of host cycles. FoldSim
exploits the static schedule to hide this latency, using different
mechanisms for the two main categories of state: target memories
and pipeline registers.

Large target memories, such as register files, typically access
only a few rows at a time. These memories use synchronous reads:
an address is presented in one target cycle, and the corresponding
data is consumed in the next. Under folding, the next evaluation
of an instance occurs after a \emph{state gap} of host cycles,
measured from the pass that issues the read. For classes without
a ready pass, this gap equals II. This gap provides slack to complete the memory access before the
instance executes again (step~6).
Golden Gate exploits similar slack to accommodate on-chip block RAM
latency~\cite{magyar2022thesis}, while Lotus stores design memories
in HBM behind caches~\cite{elsabbagh2026lotus}. FoldSim extends
this approach across the class hierarchy. When the available
state gap accommodates the HBM round trip, the requested data
arrives before its next use, allowing $H$ to remain at II.

Unlike target memories, pipeline registers are read and written
whenever their instance executes. FoldSim moves these registers
from LUTs into block RAM beside the shared logic copy, trading
greater storage capacity for narrower memory ports. To accommodate
these ports, the register state of each instance is transferred
in \emph{pages} over multiple host cycles. The next instance's
state is fetched while the current instance executes, and its
updated state is written back after its state pass
(step~1 and Fig.~\ref{fig:statev}b).

For a logic copy whose instances collectively require $B$ bits
of paged state, transferred through $P$ lanes of 72 bits each,
the resulting bandwidth constraint is
$\text{II} \ge B/(72P)$, in addition to the scheduling bound
of Eq.~\ref{eq:ii}. If the paged state exceeds the available
on-chip RAM capacity, FoldSim stores the pages in HBM instead,
transferring them through the same adapters, switches, and VM
controller used for target memories.

\subsubsection{When state is late}
Although the compiler determines when memory requests are issued,
their response times cannot always be guaranteed. To handle
variable latency, each sequencer receives an additional
\emph{advance} input (step~5). If the data required by the next
scheduled position has not yet arrived, the sequencer holds its
position, inserting an idle host cycle called a \emph{slot bubble}
(Fig.~\ref{fig:statev}c). The same mechanism applies when
the local switch lacks sufficient buffer space for the next
position's requests.

During a bubble, no new evaluation is issued, while previously
issued evaluations are allowed to drain. Consequently, the
relative distances between scheduled operations can only increase, preserving their dependencies and preventing the target from
observing host-side delays. If slot bubbles account for a fraction
$b$ of host cycles, the effective host cycles per target cycle
become $H = \text{II}/(1-b)$.

\subsection{Configurable memory modeling}
\label{sec:mem}

\subsubsection{Endpoint}
The target's memory timing must come from a model of its DRAM, not from
how long the FPGA's own HBM happens to take. Following
FASED~\cite{biancolin2019fased}, \fs{} therefore separates data storage
from memory timing. A memory endpoint stores the target's data in the
FPGA's HBM, while a timing model determines the simulated target cycle at
which each memory response should become visible. If the data from the
HBM has not arrived by that cycle, the endpoint pauses the simulator for
host cycles without advancing target time, so the target sees exactly the
modeled latency.

The memory endpoint is the one place where \fs{} still uses tokens at
runtime. Each target cycle, the simulator sends it the target's memory
outputs and receives the next cycle's inputs, a round trip of $r$ host
cycles. The compiler schedules each output early enough that its slot $s$
satisfies $s + r \le \text{II}$. When II is shorter than $r$, the endpoint
prepares the next inputs for both cases: the target accepted the pending
read response, or it did not. The simulator selects the right one at the
next frame boundary, so endpoint communication overlaps subsequent frames.

\begin{figure}[t]
  \centering
  \includegraphics[width=0.92\columnwidth]{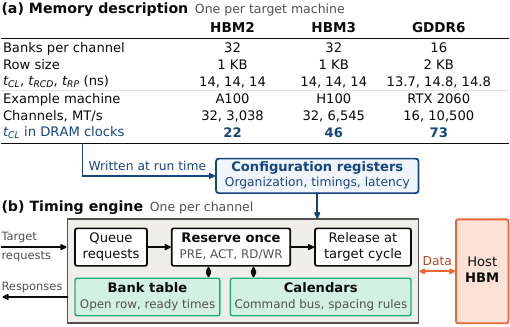}
  \caption{The configurable memory model. %
  }
  \label{fig:memsched}\label{fig:memflow}
\end{figure}

\begin{figure*}[t]
  \centering
  \includegraphics[width=\textwidth]{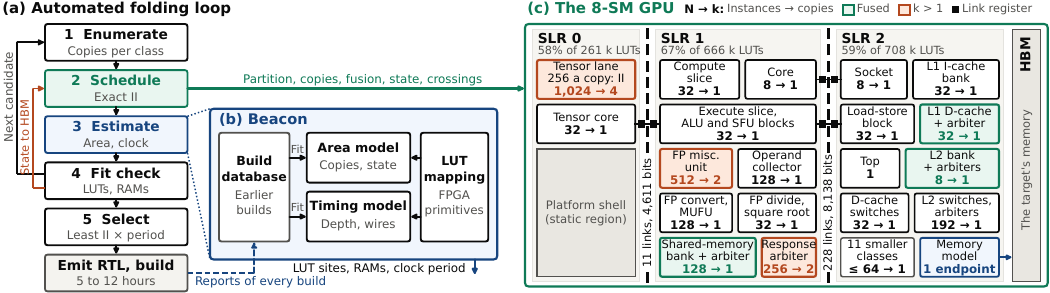}
  \caption{(a) Automated folding loop that selects the ideal number of copies of every
  class pre-synthesis using the (b) Beacon LUT
  mapping~\cite{mishchenko2007priority,mishchenko2011sop} and area-timing models~\cite{landman1971rent,donath1979placement} fitted on earlier builds realizing an (c) 8-SM GPU schedule on the FPGA.}
  \label{fig:select}\label{fig:autoloop}
\end{figure*}

\subsubsection{Memory description}
The timing engine is configured from a description
of the target memory (Fig.~\ref{fig:memflow}a).
The description specifies the DRAM standard
(HBM2, HBM3, or GDDR6), its bank organization and
timing parameters, and the target machine's channel
count, data rate, and fixed latency.

The channel count determines the number of physical
timing engines instantiated on the FPGA. The remaining
parameters are written to configuration registers
before the target resets, with timing values converted
from nanoseconds to DRAM clocks at the selected data
rate. This allows the same FPGA image to model
different memory timings without resynthesis, as
long as the channel count remains unchanged. For
example, a $t_{CL}$ of 14\,ns corresponds to 22
DRAM clocks at the A100's 3,038\,MT/s, but 46
clocks at the H100's 6,545\,MT/s.

\subsubsection{Reservation-based timing engine}
Rather than advancing a DRAM controller one clock at a time, \fs{}
schedules all the commands of a memory request once, when the request is
accepted (Fig.~\ref{fig:memsched}b). The engine tracks the open row and
ready time of each bank. It also keeps calendars of the command bus and of
the timing constraints over the next 512 DRAM clocks. For each request, it
reserves the precharge, activate, and read or write commands at their
earliest legal times. Commands to different banks can overlap, and
requests may complete out of order. Its work is proportional to the number
of requests, not to the number of DRAM clocks simulated, and each request
takes bounded work because the calendars cover a fixed window.

DRAM limits how often rows may be activated: two activates must be at
least $t_{RRD}$ apart, and at most four may fall within any window of
$t_{FAW}$. The engine enforces only the first rule. Spacing activates by
$t_{RRD}$ places the fifth at least $4t_{RRD}$ after the first, so the
second rule also holds whenever $4t_{RRD} \ge t_{FAW}$, as in the evaluated
HBM2 and HBM3 configurations. For GDDR6, where it does not, the engine
conservatively raises the effective $t_{RRD}$ to at least
$\lceil t_{FAW}/4 \rceil$.

We checked the RTL engine against a software implementation of the same
algorithm. The model has two simplifications: it serves the requests to a
bank in order, without moving row hits ahead, and it disables refresh by
default. GDDR6 timing has been tested in simulation and synthesis but is
not yet enabled through the deployed FPGA runtime.
Sec.~\ref{sec:memeval} evaluates the model's hardware cost and accuracy.

\subsection{Automated folding}
\label{sec:flow}

\subsubsection{Compiler}
Folding a design hierarchically involves three choices: how many logic
copies each class gets, whether each store of state stays on chip or goes
to HBM, and how the copies are mapped onto the FPGA's dies. Each choice
trades resources for throughput, and evaluating one by synthesis takes
hours. \fs{} therefore makes them before synthesis, in a loop around the
scheduler and an estimator (Fig.~\ref{fig:autoloop}a).

The compiler enumerates the copy counts that change how many instances a
copy serves (step 1). It runs each candidate through the scheduling algorithm
(Sec.~\ref{sec:static}), so its II is exact, and partitions it across the dies
(Sec.~\ref{sec:partition}) (step 2). Beacon (Fig.~\ref{fig:autoloop}b) then
estimates its resources and clock frequency $f$ (step 3). Beacon maps each
class to LUTs, carry chains, DSPs and RAMs with CIRCT's synthesis
passes~\cite{circt,mishchenko2007priority,mishchenko2011sop}, scales the
result by the copies and the storage chosen for each state array, and adds
wire costs from fanout, span and die
crossings~\cite{landman1971rent,donath1979placement} to the logic depth.
Beacon's area and timing models are calibrated on earlier builds. Unlike learned
predictors~\cite{xu2022sns,fang2023masterrtl,fang2024rtltimer,dai2018hlsqor,makrani2019pyramid,barboza2019preroute,guo2022timinggcn}
and autotuners~\cite{kapre2015intime,xu2017datuner}, which estimate or
search whole designs, Beacon only ranks folds of one design on one device,
where the compiler already knows the exact schedule of each candidate.

The compiler rejects candidates above 90\% of the device's LUTs or RAMs
(step 4), moving eligible state to HBM and rescheduling when on-chip state
exceeds its budget. It builds only the candidate with the lowest predicted
$\text{II}/f$ (step 5). This objective omits the slot bubbles of delayed
state accesses, which the measured $H/f$ includes.
Fig.~\ref{fig:autoloop}c shows the compiler's output for the 8-SM GPU, and
the rest of this section follows it.

\subsubsection{Area optimization}
Folding saves replicated logic but introduces per-class overhead from
sequencers, state storage, and inter-class links. For small, tightly
coupled classes this overhead is large. \emph{Class fusion} combines two
classes of one parent whose instances correspond one to one and exchange
values combinationally in both directions. The pair shares one shell, with
no link or scheduling boundary between them. In the 8-SM GPU (green boxes
in Fig.~\ref{fig:autoloop}c), each fused pair is a memory bank and its
request arbiter, in the L2, the L1 data cache and the shared memory.

A second source of area savings is the placement of instance state. The
compiler picks a storage form for each state array: flip-flops, LUT RAM,
block RAM, UltraRAM, or HBM through the virtualization mechanism of
Sec.~\ref{sec:virt}, and reschedules when HBM adds access constraints. The
8-SM GPU keeps all 95\,Mbit of its state on chip; the 4-SM GPU puts its L2
data, 18\% of its state, in HBM, and the 1,024-node datacenter about 99\%.

\subsubsection{Performance optimization}
The number of physical copies determines how much work each copy must
perform within a frame. Adding copies can reduce the resource-constrained
lower bound on II (Eq.~\ref{eq:ii}), but also increases area and may
affect clock frequency. The most useful replication is therefore in classes
that constrain the current schedule, rather than uniformly increasing the
copy count across the hierarchy.

In the 8-SM GPU (Fig.~\ref{fig:autoloop}c), four copies serve 1,024 tensor
lanes, 256 each, which sets II to 256; the FP miscellaneous unit and the
shared-memory response arbiter have two copies, and every other class has
one. These configurations illustrate how independent copy counts let \fs{}
allocate logic where it reduces the scheduling cost. For the larger
designs, the copy counts are specified in build recipes and the compiler
constructs their schedules and physical partitions. Automatic copy-count
selection is evaluated separately in Sec.~\ref{sec:autoeval}.

\subsubsection{Partitioning}
\label{sec:partition}
A configuration that fits within the total FPGA resource budget may still
be too large for any single physical region. The V80 contains three super
logic regions (SLRs), each with its own resource capacity. Neighboring
SLRs share 18,870 crossing wires, of which the compiler allows the folded
design at most 85\%. A design larger than one SLR must therefore be split,
and the split decides which links between classes cross between dies.

The compiler assigns the class copies to SLRs, one group per SLR, so that
each group fits its SLR and the links between groups fit the crossing
wires. A crossing link is registered on both sides, and its reader is
scheduled at least one host cycle after its producer. This adds no target
cycles but can lengthen the frame, so the compiler keeps an assignment only
if its schedule still reaches the initiation interval of the unpartitioned
design. Each group is emitted as a module with a floorplan constraint.

The 8-SM GPU (Fig.~\ref{fig:autoloop}c), at about 930\,k LUTs, exceeds any
single SLR. Its partitioned schedule keeps II 256, with 237 of its 3,027
links (6\% of the link bits) crossing between SLRs. The groups follow the
GPU hierarchy: the socket, caches, L2, load-store logic and memory endpoint
in SLR\,2, next to the HBM; the cores, compute slices and shared memory in
SLR\,1; and the tensor cores and lanes in SLR\,0. They use 59\%,
67\% and 58\% of the available LUTs in SLR\,2, 1 and 0, respectively.

TAPA and RapidStream~\cite{guo2023tapa,guo2022rapidstream} also register
partition boundaries, for HLS designs; \fs{} absorbs the boundary latency
into its static schedule, so the target's cycle behavior is unchanged.

\begin{figure*}[t]
  \centering
  \includegraphics[width=\textwidth]{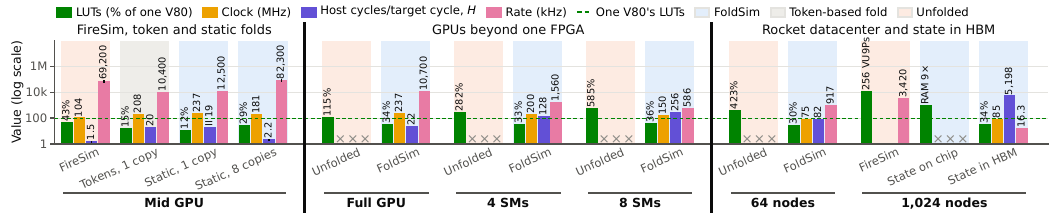}
  \caption{An overview of design points. Large Unfolded designs can't fit. FireSim's datacenter is its
  published result~\cite{karandikar2018firesim}.}
  \label{fig:showcase}
\end{figure*}

\section{Evaluation}
\label{sec:eval}

We evaluate FoldSim's ability to scale cycle-exact RTL simulation beyond single-FPGA capacity limits while maintaining practical performance. Our experiments quantify the benefits of hierarchical folding and static scheduling, examine the costs of state virtualization and the accuracy of configurable memory modeling, and assess the effectiveness of automated folding.

\subsection{Experimental setup}
\label{sec:method}\label{sec:impl}\label{sec:exact}

\begin{table}[h]
  \caption{Simulated targets: (a) key parameters; (b) unfolded LUT
  utilization against one V80.}
  \label{tab:config}
  \centering\footnotesize
  \setlength{\tabcolsep}{3pt}
  \renewcommand{\arraystretch}{1.0}
  \begin{tabular}{@{}>{\raggedright\arraybackslash}p{50pt}>{\raggedright\arraybackslash}p{182pt}@{}}
    \toprule
    \multicolumn{2}{@{}l}{(a) \textbf{GPU:} Vortex~\cite{tine2021vortex}, the SM, close to the one in A100~\cite{nvidia2020ampere}} \\
    \midrule
    ISA & RV32IMAF + virtual memory \\
    Slices, lanes & 4 per SM; 16 integer, 16 FP32, 8 load-store, 32 tensor core \\
    L1 per SM & 128\,KiB I-cache, 128\,KiB D-cache, 64\,KiB shared memory \\
    L2 & 256\,KiB and one bank per SM \\
    Memory & HBM2, one channel per SM at 900\,MT/s \\
    \midrule
    \multicolumn{2}{@{}l}{\textbf{CPU:} Rocket~\cite{asanovic2016rocket}
      datacenter, FireSim's topology~\cite{karandikar2018firesim}} \\
    \midrule
    Node & 4 Rocket cores, 16\,KiB L1 instruction and data
      caches, 256\,KiB L2, a NIC, 1\,MiB of DRAM \\
    Network & 32 top-of-rack and 4 aggregation switches, a root; 2\us{} links \\
    \bottomrule
  \end{tabular}\\[5pt]
  \setlength{\tabcolsep}{0pt}
  \begin{tabular*}{\columnwidth}{@{\extracolsep{\fill}}lcccrr@{}}
    \toprule
    (b) Without folding & Cores & Warps $\times$ threads & & LUTs & Of one V80 \\
    \midrule
    Small GPU & 2 & 8 $\times$ 4 & & 644\,k & 25\% \\
    Mid GPU & 4 & 8 $\times$ 4 & & 1,113\,k & 43\% \\
    Full GPU & 4 & 32 $\times$ 16 & & $>$\,2,956\,k & $>$\,115\% \\
    One SM & 1 & 64 $\times$ 32 & & 1,815\,k & 71\% \\
    4-SM GPU & 4 & 64 $\times$ 32 & & 7,262\,k & 282\% \\
    8-SM GPU & 8 & 64 $\times$ 32 & & 15,059\,k & 585\% \\
    \midrule
    Rocket node & 4 & & & 170\,k & 7\% \\
    64 nodes & 256 & & & 10,896\,k & 423\% \\
    1,024 nodes & 4,096 & & & 174,329\,k & 6,772\% \\
    \bottomrule
  \end{tabular*}
\end{table}

Table~\ref{tab:config} summarizes the targets and their estimated resource
requirements without folding. We use Vortex~\cite{tine2021vortex}, an
open-source RISC-V GPU, at several scales. The small and mid configurations allow comparison with unfolded FireSim; the 4-
and 8-SM configurations test designs that exceed single-FPGA capacity, which use A100-like SM resources~\cite{nvidia2020ampere}.
We also simulate a Rocket-based datacenter with FireSim's network
topology~\cite{karandikar2018firesim}, smaller nodes, and a bare-metal
network test. All designs target an AMD Alveo V80 with 2.57\,M LUTs and
32\,GB HBM, using Vivado 2025.2, CIRCT~\cite{circt,lattner2021mlir}, and SLASH, AMD Research's open-source FPGA shell and runtime~\cite{slash}. The host is a 24-core AMD Threadripper 9960X system with 256\,GB RAM.

We report time per simulated target cycle as $H/f$, where $H$ is the
average host cycles per target cycle and $f$ is the FPGA clock. The full,
4-SM, and 8-SM GPUs and both datacenters are measured on the FPGA.
For the small and mid GPUs and controlled comparisons between folding
mechanisms, we combine $H$ measured in Verilator with the routed FPGA
clock. The unfolded baseline is FireSim ported to the V80; our runtime-token baseline uses our own dynamic folding implementation. On the FPGA at 150 MHz, the 8-SM image matches the unfolded reference's instruction and cycle counts in 200 of 200 runs of both sgemm and sgemm\_tcu.

\begin{table}[h]
  \caption{Validation set across configurations.
Comparison references, workloads, and Verilator/FPGA
validation coverage. %
}
  \label{tab:designs}
  \centering\footnotesize
  \setlength{\tabcolsep}{1.5pt}
  \begin{tabular*}{\columnwidth}{@{\extracolsep{\fill}}llll@{}}
    \toprule
    Target & Baseline & In Verilator & On the FPGA \\
    \midrule
    Small GPU & Unfolded & 11 applications & \\
    Mid GPU   & Unfolded & 11 applications & \\
    Full GPU  & Four folds & demo, sgemm & demo, sgemm \\
    4 SMs & Unfolded & tcu $128^3$, demo, sgemm & tcu $128^3$, $1024^3$, demo \\
    8 SMs & Unfolded & & sgemm, tcu $128^3$ \\
    64 nodes & Unfolded & Network test & Network test \\
    1,024 nodes & Each other & Network test & Network test \\
    \bottomrule
  \end{tabular*}
\end{table}

We evaluate simulation performance and cycle accuracy using eleven Vortex applications and a datacenter network test. Table~\ref{tab:designs} summarizes the validation references and workloads. Figure~\ref{fig:showcase} provides an overview of FPGA resource usage and simulation performance across design scales.

\subsection{Hierarchical folding}
\label{sec:beyond}

Figure~\ref{fig:showcase} shows that five targets
exceed the V80's LUT capacity without folding. All fit on one V80 with routed kernels using 30--36\% of its LUTs (Tables~\ref{tab:config}
and~\ref{tab:resources}). The 8-SM GPU uses 930\,k LUTs versus an estimated 15.1\,M unfolded (16$\times$ reduction), while the
1,024-node datacenter uses 865\,k versus 174\,M
(200$\times$ reduction). The 4- and 8-SM GPUs achieve 0.641\us{} and
1.71\us{} per target cycle on the FPGA.

\begin{table}[h]
  \caption{Routed kernels of five images that run on the FPGA, and their
  build times. LUTs used as memory in parentheses.}
  \label{tab:resources}
  \centering\footnotesize
  \setlength{\tabcolsep}{1.5pt}
  \begin{tabular*}{\columnwidth}{@{\extracolsep{\fill}}lrrrrrr@{}}
    \toprule
    Image & LUTs (memory) & FFs & BRAM & URAM & DSP & Build \\
    \midrule
    V80, available & 2.57\,M & 5.15\,M & 3,741 & 1,925 & 10,848 & \\
    \midrule
    Full GPU & 863\,k (312\,k) & 1.02\,M & 425 & 64 & 209 & 12.0\,h \\
    4-SM GPU & 856\,k (298\,k) & 916\,k & 1,038 & 0 & 246 & 5.3\,h \\
    8-SM GPU & 930\,k (358\,k) & 952\,k & 716 & 305 & 246 & 5.8\,h \\
    64 nodes & 761\,k (256\,k) & 612\,k & 502 & 23 & 104 & 4.7\,h \\
    1,024 nodes & 865\,k (72\,k) & 532\,k & 1,005 & 782 & 32 & 5.8\,h \\
    \bottomrule
  \end{tabular*}
\end{table}

\begin{figure}[h]
  \centering
  \includegraphics[width=\columnwidth]{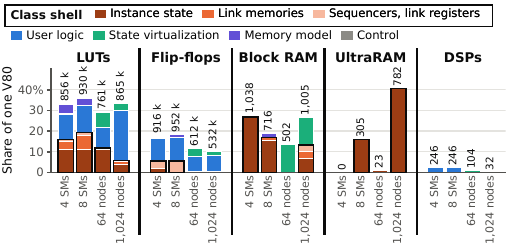}
  \caption{Four routed designs by part, as shares of one V80's
  resources; totals over the bars, the class shells outlined.}
  \label{fig:breakdown}
\end{figure}

Figure~\ref{fig:breakdown} accounts for the hardware behind these
results. Besides the shared user logic, each routed kernel includes
per-instance state, inter-class links, sequencers, virtualization hardware,
and target-memory endpoints. The class shells alone account for a
substantial share of resources: state stores, link memories, and schedule
control remain even when logic is shared. At 64 and 1,024 nodes,
virtualization hardware accounts for 26\% and 11\% of the LUTs utilization, respectively. 
Folding the 8-SM GPU's SMs (8 to 1) still leaves one full SM, 1.82 M LUTs or 71\% of the V80 (Table~\ref{tab:config}), before the shell and the memory model. Reaching 36\% takes 38 classes folded across six levels.

\begin{figure}[h]
  \centering
  \includegraphics[width=\columnwidth]{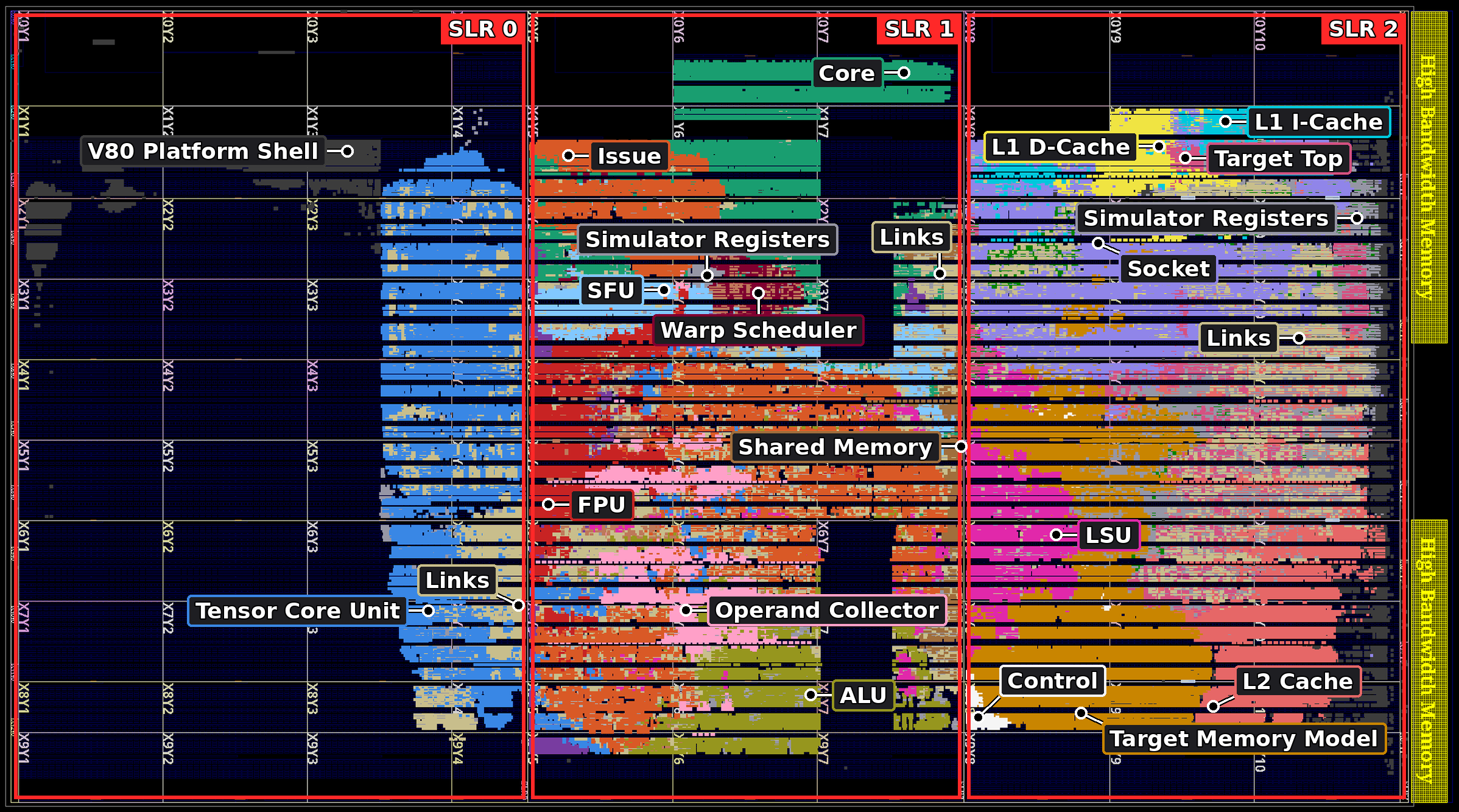}
  \caption{The routed 8-SM image on the V80; each cell is colored by the
  part of the fold it belongs to.}
  \label{fig:placement}
\end{figure}

The 8-SM FPGA placement in Fig.~\ref{fig:placement} makes this
physical organization concrete. Its folded classes occupy three SLRs:
cache and memory-facing logic lie near HBM, with compute units and tensor
lanes distributed across the remaining regions. Inter-class link storage
connects the separated copies. The 8-SM GPU is able to be routed across three SLRs, with cross-SLR links registered without increasing the scheduled II of 256.

\begin{figure}[h]
  \centering
  \includegraphics[width=\columnwidth]{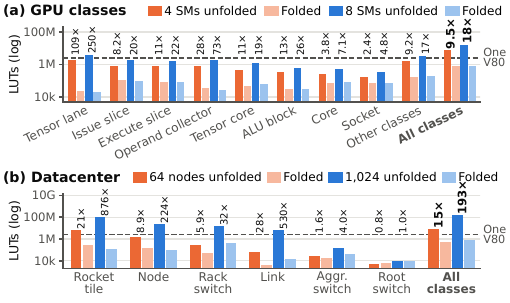}
  \caption{Per-class resources of unfolded versus folded.}
  \label{fig:classes}
\end{figure}

Figure~\ref{fig:classes} breaks down the resource savings
from hierarchical folding. For GPUs (Fig.~\ref{fig:classes}(a)),
FoldSim shares logic across sockets, cores, slices, and
execution lanes. As the design grows from 4 to 8 SMs,
the estimated unfolded logic increases from 7.26\,M
to 15.1\,M LUTs, while the folded class-level footprint
grows only from 764\,k to 853\,k, increasing the area
reduction from 9.5$\times$ to 18$\times$.
The Rocket datacenter shows similar scaling
(Fig.~\ref{fig:classes}(b)): from 64 to 1,024 nodes,
the reduction increases from 15$\times$ to 193$\times$,
with 4,096 logical Rocket tiles sharing just two
physical copies.

\begin{figure*}[t]
  \centering
  \includegraphics[width=\textwidth]{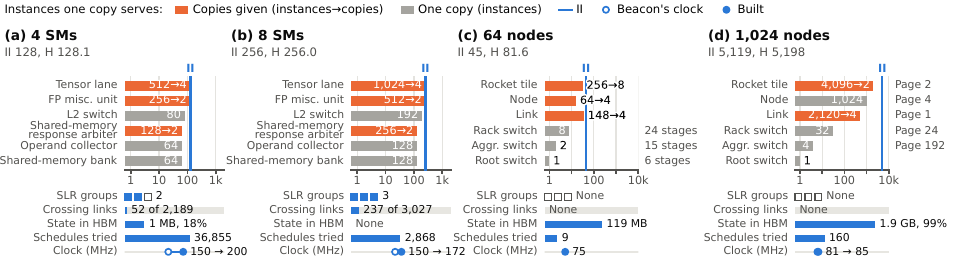}
  \caption{Four folded designs as compiled: the instances one copy serves, by class,
  against the frame (II); the copies (orange) were given in the recipes.}
  \label{fig:tradeoff}
\end{figure*}

\subsection{Static scheduling against token-based folds}
\label{sec:dynstat}\label{sec:firesim}

\begin{figure}[h]
  \centering
  \includegraphics[width=\columnwidth]{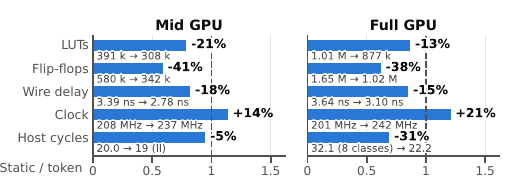}
  \caption{A static fold over a token-based fold.
  }
  \label{fig:staticeval}
\end{figure}

Figure~\ref{fig:staticeval} compares token-based and
statically scheduled folds with identical class and
copy configurations. On the mid and full GPUs, static
scheduling reduces LUTs by 21\% and 13\%, while
improving routed clock frequency by 14\% and 21\%,
respectively. Flip-flop usage also drops by 38--41\%. On the mid GPU, the matched static fold schedules II=19, against the token fold's measured $H=20$. On the full GPU, $H$ decreases from
32.1 to 22.2, although the two folds use different
class configurations.

\begin{figure}[h]
  \centering
  \includegraphics[width=\columnwidth]{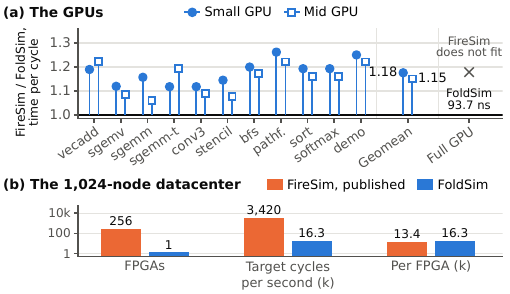}
  \caption{
  (a) FireSim's time per target cycle over
  \fs{}'s. (b) The 1,024-node datacenter against
  FireSim's data.}
  \label{fig:nsapps}
\end{figure}

Figure~\ref{fig:nsapps} compares FoldSim with FireSim
across GPUs and datacenters. Across 11 applications,
FoldSim achieves 1.12--1.26$\times$ and
1.06--1.22$\times$ FireSim's throughput on the small
and mid GPUs, with 16\% and 32\% fewer LUTs,
respectively (Fig.~\ref{fig:nsapps}(a)).
On the full GPU, FireSim cannot fit on one V80,
FoldSim achieves 93.7\ns{} per target cycle.
A larger 4-SM GPU completes a $1024\times1024$
tensor GEMM in 4.24\,s, versus 26 hours in software
RTL simulation. At datacenter scale
(Fig.~\ref{fig:nsapps}(b)), FoldSim simulates 1,024
nodes at 16.3\,k target cycles/s on one V80, compared
with FireSim's published 3.42\,M cycles/s across
256 FPGAs, yielding approximately 1.22$\times$
the throughput per FPGA.

\subsection{State virtualization}
\label{sec:virteval}\label{sec:limitsscale}

\begin{figure}[h]
  \centering
  \includegraphics[width=\columnwidth]{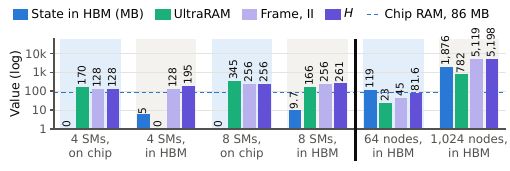}
  \caption{State on the chip and in HBM.}
  \label{fig:virtsweep}
\end{figure}

Figure~\ref{fig:virtsweep} shows the resource and
performance tradeoff of state virtualization.
Moving most of the 4-SM GPU's state (5.0 MB) to HBM saves 170 UltraRAMs but raises H from 128 to 195; backing only the L2 data array, as the routed image does, keeps $H$ at 128.1. On the 8-SM GPU, offloading 9.7\,MB
saves 179 UltraRAMs, with $H$ increasing only
from 256 to 261. The longer schedule better
hides HBM latency through prefetching.

The datacenter designs demonstrate virtualization at a
much larger scale, with 119\,MB and 1.9\,GB of HBM-backed
state at 64 and 1,024 nodes, respectively, exceeding the
V80's 86\,MB on-chip BRAM/URAM capacity.
At 64 nodes, $H=81.6$ versus II=45, while at
1,024 nodes, $H=5,198$ is only 1.5\% above II=5,119.

\subsection{Configurable memory model}
\label{sec:memeval}

\begin{table}[t]
  \caption{Memory model accuracy: target cycles in SimX against Ramulator
  (\%), by data rate (MT/s). Instructions and requests in thousands,
  writes (Wr.) in \% of requests.}
  \label{tab:memeval}
  \centering\footnotesize
  \setlength{\tabcolsep}{0pt}
  \setlength{\aboverulesep}{0pt}\setlength{\belowrulesep}{0pt}
  \newcommand{\up}{\rule{0pt}{2.3ex}}\newcommand{\dn}{\rule[-0.85ex]{0pt}{0pt}}
  \begin{tabular*}{\columnwidth}{@{\extracolsep{\fill}}lrrr@{\hspace{2.5pt}}|@{\hspace{1.5pt}}rrr@{\hspace{2.5pt}}|@{\hspace{1.5pt}}r@{\hspace{2.5pt}}|@{\hspace{1.5pt}}r@{}}
    \toprule
    \multicolumn{4}{@{}r@{\hspace{2.5pt}}|@{\hspace{1.5pt}}}{\up\dn Memory standard\hspace{4pt}} & \multicolumn{3}{c@{\hspace{2.5pt}}|@{\hspace{1.5pt}}}{HBM2} & HBM3 & GDDR6 \\
    \multicolumn{4}{@{}r@{\hspace{2.5pt}}|@{\hspace{1.5pt}}}{\up\dn Data rate (MT/s)\hspace{4pt}} & 900 & 1,752 & 3,038 & 6,545 & 15,000 \\
    \midrule
    \up\dn Application & Instr. & Req. & Wr. & \multicolumn{5}{c@{}}{Difference (\%)} \\
    \midrule
    \up vecadd, $2^{20}$ & 7,602 & 656 & 80 & $+2.52$ & $-0.66$ & $0.00$ & $0.00$ & $-1.06$ \\
    sgemv & 272 & 18.5 & 6 & $-0.21$ & $+0.61$ & $-0.83$ & $+0.23$ & $-2.69$ \\
    sgemm, 128 & 2,494 & 150 & 5 & $-0.14$ & $-0.22$ & $-0.02$ & $-0.03$ & $-0.14$ \\
    sgemm, 256 & 19,153 & 1,173 & 3 & $-0.48$ & $-0.59$ & $-0.01$ & $+0.05$ & $-0.08$ \\
    sgemm\_tcu & 47 & 2.8 & 82 & $-1.41$ & $+0.39$ & $-1.97$ & $-0.44$ & $+3.52$ \\
    conv3, 256 & 1,327 & 41.2 & 80 & $+3.19$ & $+0.22$ & $+0.44$ & $+0.15$ & $+0.21$ \\
    stencil3d & 1,786 & 53.9 & 91 & $+0.96$ & $+0.38$ & $+0.20$ & $+0.34$ & $-0.19$ \\
    bfs & 3.6 & 0.6 & 54 & $+1.33$ & $+2.18$ & $+0.71$ & $+1.75$ & $-2.39$ \\
    pathf., 512 & 3,158 & 185 & 71 & $+1.19$ & $+2.16$ & $+1.39$ & $+2.76$ & $-0.38$ \\
    sort & 724 & 0.6 & 88 & $0.00$ & $0.00$ & $0.00$ & $0.00$ & $0.00$ \\
    \dn softmax, 128 & 7,105 & 796 & 100 & $+0.05$ & $+0.03$ & $-0.01$ & $+0.01$ & $+0.30$ \\
    \midrule
    \multicolumn{4}{@{}l@{\hspace{2.5pt}}|@{\hspace{1.5pt}}}{\up\dn Mean magnitude} & 1.04 & 0.67 & 0.51 & 0.52 & 1.00 \\
    \bottomrule
  \end{tabular*}
\end{table}

Table~\ref{tab:memeval} compares FoldSim's DRAM timing model
with Ramulator under the same SimX setup. The table reports
instruction counts, memory requests, and write fractions
for 11 applications, followed by target-cycle differences
across five HBM2, HBM3, and GDDR6 configurations.
Of the 55 comparisons, 40 differ by less than 1\%,
with mean absolute errors of 0.51--1.04\% across
configurations. The largest deviation is 3.52\% for
\texttt{sgemm\_tcu} under GDDR6.

Figure~\ref{fig:breakdown} shows the memory model's
resource cost in complete routed GPU simulators.
Four memory endpoints consume 119\,k LUTs (14\% of the
kernel) on the 4-SM GPU, while a shared endpoint uses
95\,k LUTs (10\%) on the 8-SM GPU.
A standalone configurable endpoint uses 31.1\,k LUTs
and routes at 317\mhz{}, with runtime timing
configurability adding 6\% LUTs over fixed parameters.

\subsection{Automated folding}
\label{sec:autoeval}

Figure~\ref{fig:tradeoff}(a)--(d) shows four large
folded designs whose schedules and physical
mappings FoldSim generates automatically.
Four tensor-lane copies serve 512 and 1,024 GPU
instances (II=128 and 256; measured $H=128.1$
and 256.0). The 64-node datacenter requires II=45
despite at most 37 instances per copy because of
inter-class dependencies; at 1,024 nodes paging
contributes to II=5,119. The GPU mappings cross
52 and 237 registered SLR links without increasing
their scheduled II, while 1,024 nodes place about
99\% of state in HBM.

For automatic configuration selection, FoldSim explores
42 mid-GPU candidates in under a minute, compared
with an estimated 158 hours to build them all.
On the 4-SM GPU, the selected configuration is
within 5\% of the fastest among five routed candidates,
while using 21\% fewer LUTs. The clock comparisons
in Fig.~\ref{fig:tradeoff}(a), (b), and (d) show
Beacon estimates of 150, 150, and 81 MHz, against
built frequencies of 200, 172, and 85 MHz.
An earlier full-GPU candidate estimated at
2.32\,M LUTs required 2.75\,M after synthesis,
underscoring that Beacon guides configuration
selection but does not replace physical
implementation for feasibility verification.

\section{Conclusion}

\fs{} scales cycle-exact FPGA simulation by sharing RTL
logic across multiple levels of the design hierarchy
and statically scheduling execution across the resulting
shared modules. State virtualization and configurable
memory modeling further extend this approach to
large designs with substantial memory requirements.
On a single AMD Alveo V80, \fs{} runs an 8-SM GPU
and a 1,024-node Rocket datacenter, with routed
implementations requiring approximately 16$\times$
and 200$\times$ fewer LUTs than their estimated
unfolded counterparts. Together, these results
demonstrate that coordinated hierarchical sharing
can extend cycle-exact RTL simulation well beyond
the logic capacity of a single FPGA.

\bibliographystyle{ACM-Reference-Format}
\bibliography{refs}

%%% -*-BibTeX-*-
%%% Do NOT edit. File created by BibTeX with style
%%% ACM-Reference-Format-Journals [18-Jan-2012].

\begin{thebibliography}{37}

%%% ====================================================================
%%% NOTE TO THE USER: you can override these defaults by providing
%%% customized versions of any of these macros before the \bibliography
%%% command.  Each of them MUST provide its own final punctuation,
%%% except for \shownote{} and \showURL{}.  The latter does
%%% not use final punctuation, in order to avoid confusing it with
%%% the Web address.
%%%
%%% To suppress output of a particular field, define its macro to expand
%%% to an empty string, or better, \unskip, like this:
%%%
%%% \newcommand{\showURL}[1]{\unskip}   % LaTeX syntax
%%%
%%% \def \showURL #1{\unskip}           % plain TeX syntax
%%%
%%% ====================================================================

\ifx \showCODEN    \undefined \def \showCODEN     #1{\unskip}     \fi
\ifx \showISBNx    \undefined \def \showISBNx     #1{\unskip}     \fi
\ifx \showISBNxiii \undefined \def \showISBNxiii  #1{\unskip}     \fi
\ifx \showISSN     \undefined \def \showISSN      #1{\unskip}     \fi
\ifx \showLCCN     \undefined \def \showLCCN      #1{\unskip}     \fi
\ifx \shownote     \undefined \def \shownote      #1{#1}          \fi
\ifx \showarticletitle \undefined \def \showarticletitle #1{#1}   \fi
\ifx \showURL      \undefined \def \showURL       {\relax}        \fi
% The following commands are used for tagged output and should be
% invisible to TeX
\providecommand\bibfield[2]{#2}
\providecommand\bibinfo[2]{#2}
\providecommand\natexlab[1]{#1}
\providecommand\showeprint[2][]{arXiv:#2}
\makeatletter
\@ifundefined{NAT@parse@date}{}{\let\NAT@parse@date@orig\NAT@parse@date}
\@ifundefined{NAT@parse@date}{}{\def\NAT@parse@date#1#2#3#4#5#6@@{\NAT@parse@date@orig#1#2#3#4#5#6@@\def\NAT@tempyear{0000}\def\NAT@tempexlab{{?}}\ifx\NAT@year\NAT@tempyear\ifx\NAT@exlab\NAT@tempexlab\def\NAT@date{[n.\,d.]}\else\edef\NAT@date{[n.\,d.]\NAT@exlab}\fi\fi}}
\makeatother

\bibitem[Asanovi{\'c} et~al\mbox{.}(2016)]%
        {asanovic2016rocket}
\bibfield{author}{\bibinfo{person}{Krste Asanovi{\'c}}, \bibinfo{person}{Rimas
  Avi{\v{z}}ienis}, \bibinfo{person}{Jonathan Bachrach}, \bibinfo{person}{Scott
  Beamer}, \bibinfo{person}{David Biancolin}, \bibinfo{person}{Christopher
  Celio}, \bibinfo{person}{Henry Cook}, \bibinfo{person}{Daniel Dabbelt},
  \bibinfo{person}{John Hauser}, \bibinfo{person}{Adam Izraelevitz},
  \bibinfo{person}{Sagar Karandikar}, \bibinfo{person}{Ben Keller},
  \bibinfo{person}{Donggyu Kim}, \bibinfo{person}{John Koenig},
  \bibinfo{person}{Yunsup Lee}, \bibinfo{person}{Eric Love},
  \bibinfo{person}{Martin Maas}, \bibinfo{person}{Albert Magyar},
  \bibinfo{person}{Howard Mao}, \bibinfo{person}{Miquel Moreto},
  \bibinfo{person}{Albert Ou}, \bibinfo{person}{David~A. Patterson},
  \bibinfo{person}{Brian Richards}, \bibinfo{person}{Colin Schmidt},
  \bibinfo{person}{Stephen Twigg}, \bibinfo{person}{Huy Vo}, {and}
  \bibinfo{person}{Andrew Waterman}.} \bibinfo{year}{2016}\natexlab{}.
\newblock \bibinfo{booktitle}{\emph{The {Rocket} Chip Generator}}.
\newblock \bibinfo{type}{{T}echnical {R}eport} UCB/EECS-2016-17.
  \bibinfo{institution}{EECS Department, University of California, Berkeley}.
\newblock


\bibitem[Barboza et~al\mbox{.}(2019)]%
        {barboza2019preroute}
\bibfield{author}{\bibinfo{person}{Erick~Carvajal Barboza},
  \bibinfo{person}{Nishchal Shukla}, \bibinfo{person}{Yiran Chen}, {and}
  \bibinfo{person}{Jiang Hu}.} \bibinfo{year}{2019}\natexlab{}.
\newblock \showarticletitle{Machine Learning-Based Pre-Routing Timing
  Prediction with Reduced Pessimism}. In \bibinfo{booktitle}{\emph{Proceedings
  of the 56th Annual Design Automation Conference (DAC)}}.
  \bibinfo{pages}{1--6}.
\newblock
\href{https://doi.org/10.1145/3316781.3317857}{doi:\nolinkurl{10.1145/3316781.3317857}}


\bibitem[Biancolin et~al\mbox{.}(2019)]%
        {biancolin2019fased}
\bibfield{author}{\bibinfo{person}{David Biancolin}, \bibinfo{person}{Sagar
  Karandikar}, \bibinfo{person}{Donggyu Kim}, \bibinfo{person}{Jack Koenig},
  \bibinfo{person}{Andrew Waterman}, \bibinfo{person}{Jonathan Bachrach}, {and}
  \bibinfo{person}{Krste Asanovi{\'c}}.} \bibinfo{year}{2019}\natexlab{}.
\newblock \showarticletitle{{FASED}: {FPGA}-Accelerated Simulation and
  Evaluation of {DRAM}}. In \bibinfo{booktitle}{\emph{Proceedings of the
  ACM/SIGDA International Symposium on Field-Programmable Gate Arrays (FPGA)}}.
  \bibinfo{pages}{330--339}.
\newblock
\href{https://doi.org/10.1145/3289602.3293894}{doi:\nolinkurl{10.1145/3289602.3293894}}


\bibitem[Chiou et~al\mbox{.}(2007)]%
        {chiou2007fast}
\bibfield{author}{\bibinfo{person}{Derek Chiou}, \bibinfo{person}{Dam Sunwoo},
  \bibinfo{person}{Joonsoo Kim}, \bibinfo{person}{Nikhil~A. Patil},
  \bibinfo{person}{William~H. Reinhart}, \bibinfo{person}{Darrel~Eric Johnson},
  \bibinfo{person}{Jebediah Keefe}, {and} \bibinfo{person}{Hari Angepat}.}
  \bibinfo{year}{2007}\natexlab{}.
\newblock \showarticletitle{{FPGA}-Accelerated Simulation Technologies
  ({FAST}): Fast, Full-System, Cycle-Accurate Simulators}. In
  \bibinfo{booktitle}{\emph{Proceedings of the 40th Annual IEEE/ACM
  International Symposium on Microarchitecture (MICRO)}}.
  \bibinfo{pages}{249--261}.
\newblock


\bibitem[Dai et~al\mbox{.}(2018)]%
        {dai2018hlsqor}
\bibfield{author}{\bibinfo{person}{Steve Dai}, \bibinfo{person}{Yuan Zhou},
  \bibinfo{person}{Hang Zhang}, \bibinfo{person}{Ecenur Ustun},
  \bibinfo{person}{Evangeline F.~Y. Young}, {and} \bibinfo{person}{Zhiru
  Zhang}.} \bibinfo{year}{2018}\natexlab{}.
\newblock \showarticletitle{Fast and Accurate Estimation of Quality of Results
  in High-Level Synthesis with Machine Learning}. In
  \bibinfo{booktitle}{\emph{IEEE 26th Annual International Symposium on
  Field-Programmable Custom Computing Machines (FCCM)}}.
  \bibinfo{publisher}{IEEE}, \bibinfo{pages}{129--132}.
\newblock
\href{https://doi.org/10.1109/FCCM.2018.00029}{doi:\nolinkurl{10.1109/FCCM.2018.00029}}


\bibitem[Donath(1979)]%
        {donath1979placement}
\bibfield{author}{\bibinfo{person}{W. Donath}.}
  \bibinfo{year}{1979}\natexlab{}.
\newblock \showarticletitle{Placement and Average Interconnection Lengths of
  Computer Logic}.
\newblock \bibinfo{journal}{\emph{IEEE Transactions on Circuits and Systems}}
  \bibinfo{volume}{26}, \bibinfo{number}{4} (\bibinfo{year}{1979}),
  \bibinfo{pages}{272--277}.
\newblock
\href{https://doi.org/10.1109/TCS.1979.1084635}{doi:\nolinkurl{10.1109/TCS.1979.1084635}}


\bibitem[Elsabbagh et~al\mbox{.}(2026)]%
        {elsabbagh2026lotus}
\bibfield{author}{\bibinfo{person}{Fares Elsabbagh}, \bibinfo{person}{Joel~S.
  Emer}, {and} \bibinfo{person}{Daniel Sanchez}.}
  \bibinfo{year}{2026}\natexlab{}.
\newblock \showarticletitle{Lotus: A Multi-{FPGA} Task Dataflow Architecture to
  Accelerate Cycle-Level Simulation}. In \bibinfo{booktitle}{\emph{Proceedings
  of the 53rd ACM/IEEE Annual International Symposium on Computer Architecture
  (ISCA)}}. \bibinfo{pages}{1783--1796}.
\newblock
\href{https://doi.org/10.1109/ISCA66397.2026.00129}{doi:\nolinkurl{10.1109/ISCA66397.2026.00129}}


\bibitem[Emami et~al\mbox{.}(2025)]%
        {emami2025parendi}
\bibfield{author}{\bibinfo{person}{Mahyar Emami}, \bibinfo{person}{Thomas
  Bourgeat}, {and} \bibinfo{person}{James~R. Larus}.}
  \bibinfo{year}{2025}\natexlab{}.
\newblock \showarticletitle{Parendi: Thousand-Way Parallel {RTL} Simulation}.
  In \bibinfo{booktitle}{\emph{Proceedings of the 30th ACM International
  Conference on Architectural Support for Programming Languages and Operating
  Systems, Volume 2 (ASPLOS)}}. \bibinfo{pages}{783--797}.
\newblock
\href{https://doi.org/10.1145/3676641.3716010}{doi:\nolinkurl{10.1145/3676641.3716010}}


\bibitem[Emami et~al\mbox{.}(2023)]%
        {emami2023manticore}
\bibfield{author}{\bibinfo{person}{Mahyar Emami}, \bibinfo{person}{Sahand
  Kashani}, \bibinfo{person}{Keisuke Kamahori},
  \bibinfo{person}{Mohammad~Sepehr Pourghannad}, \bibinfo{person}{Ritik Raj},
  {and} \bibinfo{person}{James~R. Larus}.} \bibinfo{year}{2023}\natexlab{}.
\newblock \showarticletitle{Manticore: Hardware-Accelerated {RTL} Simulation
  with Static Bulk-Synchronous Parallelism}. In
  \bibinfo{booktitle}{\emph{Proceedings of the 28th ACM International
  Conference on Architectural Support for Programming Languages and Operating
  Systems, Volume 4 (ASPLOS)}}. \bibinfo{pages}{219--237}.
\newblock
\href{https://doi.org/10.1145/3623278.3624750}{doi:\nolinkurl{10.1145/3623278.3624750}}


\bibitem[Fang et~al\mbox{.}(2024)]%
        {fang2024rtltimer}
\bibfield{author}{\bibinfo{person}{Wenji Fang}, \bibinfo{person}{Shang Liu},
  \bibinfo{person}{Hongce Zhang}, {and} \bibinfo{person}{Zhiyao Xie}.}
  \bibinfo{year}{2024}\natexlab{}.
\newblock \showarticletitle{Annotating Slack Directly on Your {Verilog}:
  Fine-Grained {RTL} Timing Evaluation for Early Optimization}. In
  \bibinfo{booktitle}{\emph{Proceedings of the 61st ACM/IEEE Design Automation
  Conference (DAC)}}. \bibinfo{publisher}{ACM}, \bibinfo{address}{New York, NY,
  USA}, \bibinfo{pages}{1--6}.
\newblock
\href{https://doi.org/10.1145/3649329.3655671}{doi:\nolinkurl{10.1145/3649329.3655671}}


\bibitem[Fang et~al\mbox{.}(2023)]%
        {fang2023masterrtl}
\bibfield{author}{\bibinfo{person}{Wenji Fang}, \bibinfo{person}{Yao Lu},
  \bibinfo{person}{Shang Liu}, \bibinfo{person}{Qijun Zhang},
  \bibinfo{person}{Ceyu Xu}, \bibinfo{person}{Lisa~Wu Wills},
  \bibinfo{person}{Hongce Zhang}, {and} \bibinfo{person}{Zhiyao Xie}.}
  \bibinfo{year}{2023}\natexlab{}.
\newblock \showarticletitle{{MasterRTL}: A Pre-Synthesis {PPA} Estimation
  Framework for Any {RTL} Design}. In \bibinfo{booktitle}{\emph{IEEE/ACM
  International Conference on Computer Aided Design (ICCAD)}}.
  \bibinfo{publisher}{IEEE}, \bibinfo{pages}{1--9}.
\newblock
\href{https://doi.org/10.1109/ICCAD57390.2023.10323951}{doi:\nolinkurl{10.1109/ICCAD57390.2023.10323951}}


\bibitem[Guo et~al\mbox{.}(2023)]%
        {guo2023tapa}
\bibfield{author}{\bibinfo{person}{Licheng Guo}, \bibinfo{person}{Yuze Chi},
  \bibinfo{person}{Jason Lau}, \bibinfo{person}{Linghao Song},
  \bibinfo{person}{Xingyu Tian}, \bibinfo{person}{Moazin Khatti},
  \bibinfo{person}{Weikang Qiao}, \bibinfo{person}{Jie Wang},
  \bibinfo{person}{Ecenur Ustun}, \bibinfo{person}{Zhenman Fang},
  \bibinfo{person}{Zhiru Zhang}, {and} \bibinfo{person}{Jason Cong}.}
  \bibinfo{year}{2023}\natexlab{}.
\newblock \showarticletitle{{TAPA}: A Scalable Task-Parallel Dataflow
  Programming Framework for Modern {FPGAs} with Co-Optimization of {HLS} and
  Physical Design}.
\newblock \bibinfo{journal}{\emph{ACM Transactions on Reconfigurable Technology
  and Systems}} \bibinfo{volume}{16}, \bibinfo{number}{4}, Article
  \bibinfo{articleno}{63} (\bibinfo{year}{2023}).
\newblock
\href{https://doi.org/10.1145/3609335}{doi:\nolinkurl{10.1145/3609335}}


\bibitem[Guo et~al\mbox{.}(2022b)]%
        {guo2022rapidstream}
\bibfield{author}{\bibinfo{person}{Licheng Guo}, \bibinfo{person}{Pongstorn
  Maidee}, \bibinfo{person}{Yun Zhou}, \bibinfo{person}{Chris Lavin},
  \bibinfo{person}{Jie Wang}, \bibinfo{person}{Yuze Chi},
  \bibinfo{person}{Weikang Qiao}, \bibinfo{person}{Alireza Kaviani},
  \bibinfo{person}{Zhiru Zhang}, {and} \bibinfo{person}{Jason Cong}.}
  \bibinfo{year}{2022}\natexlab{b}.
\newblock \showarticletitle{{RapidStream}: Parallel Physical Implementation of
  {FPGA} {HLS} Designs}. In \bibinfo{booktitle}{\emph{Proceedings of the 2022
  ACM/SIGDA International Symposium on Field-Programmable Gate Arrays (FPGA)}}.
  \bibinfo{pages}{1--12}.
\newblock
\href{https://doi.org/10.1145/3490422.3502361}{doi:\nolinkurl{10.1145/3490422.3502361}}


\bibitem[Guo et~al\mbox{.}(2022a)]%
        {guo2022timinggcn}
\bibfield{author}{\bibinfo{person}{Zizheng Guo}, \bibinfo{person}{Mingjie Liu},
  \bibinfo{person}{Jiaqi Gu}, \bibinfo{person}{Shuhan Zhang},
  \bibinfo{person}{David~Z. Pan}, {and} \bibinfo{person}{Yibo Lin}.}
  \bibinfo{year}{2022}\natexlab{a}.
\newblock \showarticletitle{A Timing Engine Inspired Graph Neural Network Model
  for Pre-Routing Slack Prediction}. In \bibinfo{booktitle}{\emph{Proceedings
  of the 59th ACM/IEEE Design Automation Conference (DAC)}}.
  \bibinfo{pages}{1207--1212}.
\newblock
\href{https://doi.org/10.1145/3489517.3530597}{doi:\nolinkurl{10.1145/3489517.3530597}}


\bibitem[Guo et~al\mbox{.}(2025)]%
        {guo2025gem}
\bibfield{author}{\bibinfo{person}{Zizheng Guo}, \bibinfo{person}{Yanqing
  Zhang}, \bibinfo{person}{Runsheng Wang}, \bibinfo{person}{Yibo Lin}, {and}
  \bibinfo{person}{Haoxing Ren}.} \bibinfo{year}{2025}\natexlab{}.
\newblock \showarticletitle{{GEM}: {GPU}-Accelerated Emulator-Inspired {RTL}
  Simulation}. In \bibinfo{booktitle}{\emph{Proceedings of the 62nd ACM/IEEE
  Design Automation Conference (DAC)}}. \bibinfo{pages}{1--7}.
\newblock
\href{https://doi.org/10.1109/DAC63849.2025.11132713}{doi:\nolinkurl{10.1109/DAC63849.2025.11132713}}


\bibitem[Kapre et~al\mbox{.}(2015)]%
        {kapre2015intime}
\bibfield{author}{\bibinfo{person}{Nachiket Kapre}, \bibinfo{person}{Harnhua
  Ng}, \bibinfo{person}{Kirvy Teo}, {and} \bibinfo{person}{Jaco Naude}.}
  \bibinfo{year}{2015}\natexlab{}.
\newblock \showarticletitle{{InTime}: A Machine Learning Approach for Efficient
  Selection of {FPGA} {CAD} Tool Parameters}. In
  \bibinfo{booktitle}{\emph{Proceedings of the ACM/SIGDA International
  Symposium on Field-Programmable Gate Arrays (FPGA)}}.
  \bibinfo{publisher}{ACM}, \bibinfo{address}{New York, NY, USA},
  \bibinfo{pages}{23--26}.
\newblock
\href{https://doi.org/10.1145/2684746.2689081}{doi:\nolinkurl{10.1145/2684746.2689081}}


\bibitem[Karandikar et~al\mbox{.}(2018)]%
        {karandikar2018firesim}
\bibfield{author}{\bibinfo{person}{Sagar Karandikar}, \bibinfo{person}{Howard
  Mao}, \bibinfo{person}{Donggyu Kim}, \bibinfo{person}{David Biancolin},
  \bibinfo{person}{Alon Amid}, \bibinfo{person}{Dayeol Lee},
  \bibinfo{person}{Nathan Pemberton}, \bibinfo{person}{Emmanuel Amaro},
  \bibinfo{person}{Colin Schmidt}, \bibinfo{person}{Aditya Chopra},
  \bibinfo{person}{Qijing Huang}, \bibinfo{person}{Kyle Kovacs},
  \bibinfo{person}{Borivoje Nikoli{\'c}}, \bibinfo{person}{Randy Katz},
  \bibinfo{person}{Jonathan Bachrach}, {and} \bibinfo{person}{Krste
  Asanovi{\'c}}.} \bibinfo{year}{2018}\natexlab{}.
\newblock \showarticletitle{{FireSim}: {FPGA}-Accelerated Cycle-Exact Scale-Out
  System Simulation in the Public Cloud}. In
  \bibinfo{booktitle}{\emph{Proceedings of the 45th Annual International
  Symposium on Computer Architecture (ISCA)}}. \bibinfo{pages}{29--42}.
\newblock
\href{https://doi.org/10.1109/ISCA.2018.00014}{doi:\nolinkurl{10.1109/ISCA.2018.00014}}


\bibitem[Landman and Russo(1971)]%
        {landman1971rent}
\bibfield{author}{\bibinfo{person}{B.~S. Landman} {and} \bibinfo{person}{R.~L.
  Russo}.} \bibinfo{year}{1971}\natexlab{}.
\newblock \showarticletitle{On a Pin Versus Block Relationship for Partitions
  of Logic Graphs}.
\newblock \bibinfo{journal}{\emph{IEEE Trans. Comput.}} \bibinfo{volume}{C-20},
  \bibinfo{number}{12} (\bibinfo{year}{1971}), \bibinfo{pages}{1469--1479}.
\newblock
\href{https://doi.org/10.1109/T-C.1971.223159}{doi:\nolinkurl{10.1109/T-C.1971.223159}}


\bibitem[Lattner et~al\mbox{.}(2021)]%
        {lattner2021mlir}
\bibfield{author}{\bibinfo{person}{Chris Lattner}, \bibinfo{person}{Mehdi
  Amini}, \bibinfo{person}{Uday Bondhugula}, \bibinfo{person}{Albert Cohen},
  \bibinfo{person}{Andy Davis}, \bibinfo{person}{Jacques Pienaar},
  \bibinfo{person}{River Riddle}, \bibinfo{person}{Tatiana Shpeisman},
  \bibinfo{person}{Nicolas Vasilache}, {and} \bibinfo{person}{Oleksandr
  Zinenko}.} \bibinfo{year}{2021}\natexlab{}.
\newblock \showarticletitle{{MLIR}: Scaling Compiler Infrastructure for Domain
  Specific Computation}. In \bibinfo{booktitle}{\emph{Proceedings of the
  IEEE/ACM International Symposium on Code Generation and Optimization (CGO)}}.
  \bibinfo{pages}{2--14}.
\newblock
\href{https://doi.org/10.1109/CGO51591.2021.9370308}{doi:\nolinkurl{10.1109/CGO51591.2021.9370308}}


\bibitem[Lin et~al\mbox{.}(2022)]%
        {lin2022rtlflow}
\bibfield{author}{\bibinfo{person}{Dian-Lun Lin}, \bibinfo{person}{Haoxing
  Ren}, \bibinfo{person}{Yanqing Zhang}, \bibinfo{person}{Brucek Khailany},
  {and} \bibinfo{person}{Tsung-Wei Huang}.} \bibinfo{year}{2022}\natexlab{}.
\newblock \showarticletitle{From {RTL} to {CUDA}: A {GPU} Acceleration Flow for
  {RTL} Simulation with Batch Stimulus}. In
  \bibinfo{booktitle}{\emph{Proceedings of the 51st International Conference on
  Parallel Processing (ICPP)}}. \bibinfo{pages}{1--12}.
\newblock
\href{https://doi.org/10.1145/3545008.3545091}{doi:\nolinkurl{10.1145/3545008.3545091}}


\bibitem[Magyar(2022)]%
        {magyar2022thesis}
\bibfield{author}{\bibinfo{person}{Albert Magyar}.}
  \bibinfo{year}{2022}\natexlab{}.
\newblock \emph{\bibinfo{title}{Improving {FPGA} Simulation Capacity with
  Automatic Resource Multi-Threading}}.
\newblock \bibinfo{thesistype}{Ph.\,D. Dissertation}.
  \bibinfo{school}{University of California, Berkeley}.
\newblock
\newblock
\shownote{Technical Report UCB/EECS-2022-24}.


\bibitem[Magyar et~al\mbox{.}(2019)]%
        {magyar2019goldengate}
\bibfield{author}{\bibinfo{person}{Albert Magyar}, \bibinfo{person}{David
  Biancolin}, \bibinfo{person}{John Koenig}, \bibinfo{person}{Sanjit Seshia},
  \bibinfo{person}{Jonathan Bachrach}, {and} \bibinfo{person}{Krste
  Asanovi{\'c}}.} \bibinfo{year}{2019}\natexlab{}.
\newblock \showarticletitle{Golden {Gate}: Bridging the Resource-Efficiency Gap
  Between {ASICs} and {FPGA} Prototypes}. In
  \bibinfo{booktitle}{\emph{Proceedings of the IEEE/ACM International
  Conference on Computer-Aided Design (ICCAD)}}. \bibinfo{pages}{1--8}.
\newblock
\href{https://doi.org/10.1109/ICCAD45719.2019.8942087}{doi:\nolinkurl{10.1109/ICCAD45719.2019.8942087}}


\bibitem[Makrani et~al\mbox{.}(2019)]%
        {makrani2019pyramid}
\bibfield{author}{\bibinfo{person}{Hosein~Mohammadi Makrani},
  \bibinfo{person}{Farnoud Farahmand}, \bibinfo{person}{Hossein Sayadi},
  \bibinfo{person}{Sara Bondi}, \bibinfo{person}{Sai Manoj~Pudukotai
  Dinakarrao}, \bibinfo{person}{Houman Homayoun}, {and}
  \bibinfo{person}{Setareh Rafatirad}.} \bibinfo{year}{2019}\natexlab{}.
\newblock \showarticletitle{{Pyramid}: Machine Learning Framework to Estimate
  the Optimal Timing and Resource Usage of a High-Level Synthesis Design}. In
  \bibinfo{booktitle}{\emph{29th International Conference on Field Programmable
  Logic and Applications (FPL)}}. \bibinfo{publisher}{IEEE},
  \bibinfo{pages}{397--403}.
\newblock
\href{https://doi.org/10.1109/FPL.2019.00069}{doi:\nolinkurl{10.1109/FPL.2019.00069}}


\bibitem[Mishchenko et~al\mbox{.}(2011)]%
        {mishchenko2011sop}
\bibfield{author}{\bibinfo{person}{Alan Mishchenko}, \bibinfo{person}{Robert
  Brayton}, \bibinfo{person}{Stephen Jang}, {and} \bibinfo{person}{Victor
  Kravets}.} \bibinfo{year}{2011}\natexlab{}.
\newblock \showarticletitle{Delay Optimization Using {SOP} Balancing}. In
  \bibinfo{booktitle}{\emph{IEEE/ACM International Conference on Computer-Aided
  Design (ICCAD)}}. \bibinfo{pages}{375--382}.
\newblock
\href{https://doi.org/10.1109/ICCAD.2011.6105357}{doi:\nolinkurl{10.1109/ICCAD.2011.6105357}}


\bibitem[Mishchenko et~al\mbox{.}(2007)]%
        {mishchenko2007priority}
\bibfield{author}{\bibinfo{person}{Alan Mishchenko}, \bibinfo{person}{Sungmin
  Cho}, \bibinfo{person}{Satrajit Chatterjee}, {and} \bibinfo{person}{Robert
  Brayton}.} \bibinfo{year}{2007}\natexlab{}.
\newblock \showarticletitle{Combinational and Sequential Mapping with Priority
  Cuts}. In \bibinfo{booktitle}{\emph{IEEE/ACM International Conference on
  Computer-Aided Design (ICCAD)}}. \bibinfo{pages}{354--361}.
\newblock
\href{https://doi.org/10.1109/ICCAD.2007.4397290}{doi:\nolinkurl{10.1109/ICCAD.2007.4397290}}


\bibitem[{NVIDIA Corporation}(2020)]%
        {nvidia2020ampere}
\bibfield{author}{\bibinfo{person}{{NVIDIA Corporation}}.}
  \bibinfo{year}{2020}\natexlab{}.
\newblock \bibinfo{title}{{NVIDIA A100 Tensor Core GPU Architecture}}.
\newblock
  \bibinfo{howpublished}{\url{https://images.nvidia.com/aem-dam/en-zz/Solutions/data-center/nvidia-ampere-architecture-whitepaper.pdf}}.
\newblock


\bibitem[Pellauer et~al\mbox{.}(2011)]%
        {pellauer2011hasim}
\bibfield{author}{\bibinfo{person}{Michael Pellauer}, \bibinfo{person}{Michael
  Adler}, \bibinfo{person}{Michel Kinsy}, \bibinfo{person}{Angshuman Parashar},
  {and} \bibinfo{person}{Joel Emer}.} \bibinfo{year}{2011}\natexlab{}.
\newblock \showarticletitle{{HAsim}: {FPGA}-Based High-Detail Multicore
  Simulation Using Time-Division Multiplexing}. In
  \bibinfo{booktitle}{\emph{Proceedings of the 17th IEEE International
  Symposium on High Performance Computer Architecture (HPCA)}}.
  \bibinfo{pages}{406--417}.
\newblock
\href{https://doi.org/10.1109/HPCA.2011.5749747}{doi:\nolinkurl{10.1109/HPCA.2011.5749747}}


\bibitem[Rau(1994)]%
        {rau1994modulo}
\bibfield{author}{\bibinfo{person}{B.~Ramakrishna Rau}.}
  \bibinfo{year}{1994}\natexlab{}.
\newblock \showarticletitle{Iterative Modulo Scheduling: An Algorithm for
  Software Pipelining Loops}. In \bibinfo{booktitle}{\emph{Proceedings of the
  27th Annual International Symposium on Microarchitecture (MICRO)}}.
  \bibinfo{pages}{63--74}.
\newblock
\href{https://doi.org/10.1145/192724.192731}{doi:\nolinkurl{10.1145/192724.192731}}


\bibitem[Tan et~al\mbox{.}(2010a)]%
        {tan2010rampgold}
\bibfield{author}{\bibinfo{person}{Zhangxi Tan}, \bibinfo{person}{Andrew
  Waterman}, \bibinfo{person}{Rimas Avizienis}, \bibinfo{person}{Yunsup Lee},
  \bibinfo{person}{Henry Cook}, \bibinfo{person}{David Patterson}, {and}
  \bibinfo{person}{Krste Asanovi{\'c}}.} \bibinfo{year}{2010}\natexlab{a}.
\newblock \showarticletitle{{RAMP} {Gold}: An {FPGA}-based Architecture
  Simulator for Multiprocessors}. In \bibinfo{booktitle}{\emph{Proceedings of
  the 47th Design Automation Conference (DAC)}}. \bibinfo{pages}{463--468}.
\newblock
\href{https://doi.org/10.1145/1837274.1837390}{doi:\nolinkurl{10.1145/1837274.1837390}}


\bibitem[Tan et~al\mbox{.}(2010b)]%
        {tan2010fame}
\bibfield{author}{\bibinfo{person}{Zhangxi Tan}, \bibinfo{person}{Andrew
  Waterman}, \bibinfo{person}{Henry Cook}, \bibinfo{person}{Sarah Bird},
  \bibinfo{person}{Krste Asanovi{\'c}}, {and} \bibinfo{person}{David
  Patterson}.} \bibinfo{year}{2010}\natexlab{b}.
\newblock \showarticletitle{A Case for {FAME}: {FPGA} Architecture Model
  Execution}. In \bibinfo{booktitle}{\emph{Proceedings of the 37th Annual
  International Symposium on Computer Architecture (ISCA)}}.
  \bibinfo{pages}{290--301}.
\newblock
\href{https://doi.org/10.1145/1815961.1815999}{doi:\nolinkurl{10.1145/1815961.1815999}}


\bibitem[{The CIRCT project}(0000)]%
        {circt}
\bibfield{author}{\bibinfo{person}{{The CIRCT project}}.}
  \bibinfo{year}{[n.\,d.]}\natexlab{}.
\newblock \bibinfo{title}{{CIRCT}: Circuit {IR} Compilers and Tools}.
\newblock \bibinfo{howpublished}{\url{https://circt.llvm.org}}.
\newblock


\bibitem[{The SLASH developers}(0000)]%
        {slash}
\bibfield{author}{\bibinfo{person}{{The SLASH developers}}.}
  \bibinfo{year}{[n.\,d.]}\natexlab{}.
\newblock \bibinfo{title}{{SLASH}: An Open-Source Research Shell for {Alveo
  V80}}.
\newblock \bibinfo{howpublished}{\url{https://slash-fpga.readthedocs.io/}}.
\newblock
\newblock
\shownote{Accessed 2026-10-27}.


\bibitem[Tine et~al\mbox{.}(2021)]%
        {tine2021vortex}
\bibfield{author}{\bibinfo{person}{Blaise Tine},
  \bibinfo{person}{Krishna~Praveen Yalamarthy}, \bibinfo{person}{Fares
  Elsabbagh}, {and} \bibinfo{person}{Hyesoon Kim}.}
  \bibinfo{year}{2021}\natexlab{}.
\newblock \showarticletitle{Vortex: Extending the {RISC-V} {ISA} for {GPGPU}
  and {3D}-Graphics}. In \bibinfo{booktitle}{\emph{Proceedings of the 54th
  Annual IEEE/ACM International Symposium on Microarchitecture (MICRO)}}.
  \bibinfo{pages}{754--766}.
\newblock
\href{https://doi.org/10.1145/3466752.3480128}{doi:\nolinkurl{10.1145/3466752.3480128}}


\bibitem[Wang and Beamer(2023)]%
        {wang2023repcut}
\bibfield{author}{\bibinfo{person}{Haoyuan Wang} {and} \bibinfo{person}{Scott
  Beamer}.} \bibinfo{year}{2023}\natexlab{}.
\newblock \showarticletitle{{RepCut}: Superlinear Parallel {RTL} Simulation
  with Replication-Aided Partitioning}. In
  \bibinfo{booktitle}{\emph{Proceedings of the 28th ACM International
  Conference on Architectural Support for Programming Languages and Operating
  Systems, Volume 3 (ASPLOS)}}. \bibinfo{pages}{572--585}.
\newblock
\href{https://doi.org/10.1145/3582016.3582034}{doi:\nolinkurl{10.1145/3582016.3582034}}


\bibitem[Whangbo et~al\mbox{.}(2024)]%
        {whangbo2024fireaxe}
\bibfield{author}{\bibinfo{person}{Joonho Whangbo}, \bibinfo{person}{Edwin
  Lim}, \bibinfo{person}{Chengyi~Lux Zhang}, \bibinfo{person}{Kevin Anderson},
  \bibinfo{person}{Abraham Gonzalez}, \bibinfo{person}{Raghav Gupta},
  \bibinfo{person}{Nivedha Krishnakumar}, \bibinfo{person}{Sagar Karandikar},
  \bibinfo{person}{Borivoje Nikoli{\'c}}, \bibinfo{person}{Yakun~Sophia Shao},
  {and} \bibinfo{person}{Krste Asanovi{\'c}}.} \bibinfo{year}{2024}\natexlab{}.
\newblock \showarticletitle{{FireAxe}: Partitioned {FPGA}-Accelerated
  Simulation of Large-Scale {RTL} Designs}. In
  \bibinfo{booktitle}{\emph{Proceedings of the 51st ACM/IEEE Annual
  International Symposium on Computer Architecture (ISCA)}}.
  \bibinfo{pages}{501--515}.
\newblock
\href{https://doi.org/10.1109/ISCA59077.2024.00044}{doi:\nolinkurl{10.1109/ISCA59077.2024.00044}}


\bibitem[Xu et~al\mbox{.}(2022)]%
        {xu2022sns}
\bibfield{author}{\bibinfo{person}{Ceyu Xu}, \bibinfo{person}{Chris
  Kjellqvist}, {and} \bibinfo{person}{Lisa~Wu Wills}.}
  \bibinfo{year}{2022}\natexlab{}.
\newblock \showarticletitle{{SNS}'s Not a Synthesizer: A Deep-Learning-Based
  Synthesis Predictor}. In \bibinfo{booktitle}{\emph{Proceedings of the 49th
  Annual International Symposium on Computer Architecture (ISCA)}}.
  \bibinfo{publisher}{ACM}, \bibinfo{address}{New York, NY, USA},
  \bibinfo{pages}{847--859}.
\newblock
\href{https://doi.org/10.1145/3470496.3527444}{doi:\nolinkurl{10.1145/3470496.3527444}}


\bibitem[Xu et~al\mbox{.}(2017)]%
        {xu2017datuner}
\bibfield{author}{\bibinfo{person}{Chang Xu}, \bibinfo{person}{Gai Liu},
  \bibinfo{person}{Ritchie Zhao}, \bibinfo{person}{Stephen Yang},
  \bibinfo{person}{Guojie Luo}, {and} \bibinfo{person}{Zhiru Zhang}.}
  \bibinfo{year}{2017}\natexlab{}.
\newblock \showarticletitle{A Parallel Bandit-Based Approach for Autotuning
  {FPGA} Compilation}. In \bibinfo{booktitle}{\emph{Proceedings of the
  ACM/SIGDA International Symposium on Field-Programmable Gate Arrays (FPGA)}}.
  \bibinfo{publisher}{ACM}, \bibinfo{address}{New York, NY, USA},
  \bibinfo{pages}{157--166}.
\newblock
\href{https://doi.org/10.1145/3020078.3021747}{doi:\nolinkurl{10.1145/3020078.3021747}}


\end{thebibliography}

\end{document}